\documentclass[numberedappendix]{emulateapj}
\usepackage{apjfonts}
\newcommand{\be}{\begin{equation}}
\newcommand{\ee}{\end{equation}}
\newcommand{\bea}{\begin{eqnarray}}
\newcommand{\eea}{\end{eqnarray}}

\newcommand{\kms}{{\,{\mathrm{km}}\,{\mathrm{s}}^{-1}}}

\shortauthors{CONROY AND OSTRIKER}
\shorttitle{Thermal Balance in the ICM}

\begin{document}

\title{Thermal Balance in the Intracluster Medium: Is AGN Feedback
  Necessary?}

\author{Charlie Conroy \& Jeremiah P. Ostriker} 
\affil{Department of Astrophysical Sciences, Princeton University,
  Princeton, NJ 08544}

\slugcomment{Submitted to ApJ, 6 Dec 2007}

\begin{abstract}

  A variety of physical heating mechanisms are combined with radiative
  cooling to explore, via one dimensional hydrodynamic simulations,
  the expected thermal properties of the intracluster medium (ICM) in
  the context of the cooling flow problem.  Energy injection from type
  Ia supernovae, thermal conduction, and dynamical friction (DF) from
  orbiting satellite galaxies are considered as potential heating
  mechanisms of the central regions of the ICM, both separately and in
  conjunction.  The novel feature of this work is the exploration of a
  wide range of efficiencies of each heating process.  While DF and
  conduction can provide a substantial amount of energy, neither
  mechanism operating alone can produce nor maintain an ICM in thermal
  balance over cosmological timescales, in stark contrast with
  observations.  For simulated clusters with initially isothermal
  temperature profiles, both mechanisms acting \emph{in combination}
  result in long-term thermal balance for a range of ICM temperatures
  and for central electron densities less than $n_e\sim 0.02$
  cm$^{-3}$; at greater densities catastrophic cooling invariably
  occurs.  Furthermore, these heating mechanisms can neither produce
  nor maintain clusters with a declining temperature profile in the
  central regions, implying that the observed ``cooling-core''
  clusters, which have such declining temperature profiles, cannot be
  maintained with these mechanisms alone.  Supernovae heating also
  fails to maintain clusters in thermal balance for cosmological
  timescales since such heating is largely unresponsive to the
  properties of the ICM.  Thus, while there appears to be an abundant
  supply of energy capable of heating the ICM in clusters, it is
  extremely difficult for the energy deposition to occur in such a way
  that the ICM remains in thermal {\it balance} over cosmological
  time-scales.  For intracluster media that are not in thermal
  balance, the addition of a small amount of relativistic pressure
  (provided by e.g. cosmic rays) forestalls neither catastrophic
  heating nor cooling.  These conclusions are driven largely by the
  fact that 1) DF heating scales approximately as the gas density,
  while cooling scales as gas density squared, and thus DF heating
  cannot generically balance cooling without fine-tuning; 2)
  conduction acts to erase temperature gradients, while most observed
  clusters in fact show strong gradients in the inner regions.  These
  results strongly suggest that a more dynamic heating process such as
  feedback from a central black hole is required to generate the
  properties of observed intracluster media.

\end{abstract}


\section{Introduction}
\label{section:intro}

Groups and clusters of galaxies are filled with hot plasma in at least
approximate pressure equilibrium with the gravitational potential of
their dark matter halo.  Observations have demonstrated for decades
that the cooling time of central regions of this intracluster medium
(ICM) in most ($>70$\%) clusters is shorter than a Hubble time
\citep[e.g.][]{Edge92, Peres98, Sanderson06, Vikhlinin06a}; indeed, it
is often as short as $\sim 0.1-1$ Gyr.  In the absence of heating, the
ICM will thus cool and flow into the central galaxy at the prodigious
rate of hundreds of solar masses per year \citep{Cowie77, Fabian77a,
  Mathews78}; see \citet{Fabian94} for a review.

Since this (catastrophically) cooling gas has never been observed
\citep{Peterson01, Peterson03, Tamura01}, it is now generally supposed
that there exists one or more heating mechanisms that maintains the
ICM in overall thermal balance.  The tension created by the fact that
the ICM is not observed to be significantly cooling, despite the short
cooling times near the cluster center, has become known as the cooling
flow problem.

Many possible heating mechanisms have been investigated, including
thermal conduction \citep[e.g.][]{Binney81, Tucker83, Voigt02,
  Fabian02, Zakamska03, Kim03a, Voigt04, Dolag04, Pope06} which
carries heat from the abundant thermal reservoir of the cluster gas to
the cooling inner parts, gravitational heating, including the heating
due to the orbital motions of galaxies, i.e. dynamical friction
\citep[DF; e.g.][]{Miller86, Just90, Fabian03, ElZant04b, Kim05,
  Dekel07}, turbulent mixing \citep[e.g.][]{Deiss96, Kim03b, Voigt04,
  Dennis05}, energy injection from active galactic nuclei (AGN) via
bubbles, sound waves, etc. \citep{Binney95, Nusser06, Binney07,
  Ruszkowski04a, Ruszkowski04b, Fujita05, Mathews06, Ciotti01, Voit05,
  Ciotti07}, and combinations of conduction and AGN
\citep{Ruszkowski02, Brighenti03, Fujita05} or conduction and
cosmic-rays \citep[CRs;][]{Guo07}, or AGN and preheating
\citep{McCarthy07c}.  One of the most comprehensive studies to date
was undertaken by \citet{Brighenti02} who included many of the heating
mechanisms mentioned above and found that no combination generically
reproduced observations.  A stringent constraint on the potential
heating mechanism is that it must not only supply of order the
necessary energy but must also supply it in a way that approximately
balances cooling (locally) so that the ICM does not either heat up or
cool on cosmological time-scales.

Many of these mechanisms have been idealized, and it is unclear how
effective they would be in observed clusters.  For example, though
difficult to constrain observationally, it has been suggested that the
conductivity in clusters may be much smaller than is required to stem
the cooling flows \citep{Markevitch03, Xiang07}.  In addition, while
DF heating may be attractive, since there are a plethora of satellite
galaxies in clusters, it has yet to be identified as important in
hydrodynamical simulations \citep{Faltenbacher05}, although this may
be due to insufficient resolution \citep{Naab07}.  Finally, bubbles
from activity in central galaxies may be important
\citep[e.g.][]{Churazov01, Ruszkowski02, Bruggen02}, but the mechanism
by which they transfer their energy to the cluster gas remains
obscure.

Even if one or more of these mechanisms could supply the requisite
energy (i.e. result in thermal balance), the mechanism must, in
addition, not allow the gas to be thermally unstable \citep{Field65,
  Balbus86}.  In what follows we make extensive use of this
distinction between thermal {\it balance}, where the net cooling of a
given parcel of gas is zero, and thermal {\it stability}, which
pertains to the ability of a gas to remain in thermal balance in the
presence of isobaric perturbations.  While attention in the literature
has focused on investigating the stability of an ICM in thermal
balance \citep[e.g.][]{Kim03b}, herein we address the more fundamental
challenge of maintaining a cluster in thermal balance with one or more
of the heating mechanisms described above.

The physical mechanism causing thermal instability is easy to
understand.  Small regions which are slightly over-dense will radiate
more than their surroundings (since cooling per unit volume scales as
$\rho^2$) and, isobarically contracting, find a new equilibrium, which
is at a higher density and lower temperature.  For the temperature
domain in question, this leads to further cooling and a thermal
runaway results.  If a heating process such as conduction or DF
heating exists, there will be a stabilizing influence.  But if the
process scales as $\rho$ (per unit volume) then the equilibrium will
not be stable, so high temperature under-dense regions will heat
exponentially and lower temperature over-dense regions will still cool
exponentially.  However, a non-thermal component (such as cosmic rays
or tangled magnetic fields) can in principle suppress this instability
\citep[][see also the appendix herein]{Cen05, Guo07} because the
non-thermal pressure partially decouples the hydrodynamic balance from
the thermodynamics of the cluster gas.  In other words, an ICM with a
non-thermal component that radiatively cools will lose less total
pressure support than an ICM with no non-thermal component.  An ICM
with a non-thermal component will thus have to contract less to
compensate for the lost thermal pressure support.

Observed clusters can be rather cleanly divided into two categories
based on the properties of their ICM.  ``Cooling core'' (CC) clusters
are those which have steep temperature drops in their central regions
\citep{Sanderson06} and metallicity gradients \citep{DeGrandi01}; many
CC clusters have identified radio emission at their centers
\citep{Best05} and show signs of AGN activity, including observed
sound waves \citep[e.g.][]{Fabian06} apparently emanating from their
center, and bubbles at an average distance of $\sim20$ kpc from the
cluster center \citep{Birzan04}.  In contrast, non-cooling core (NCC)
clusters are approximately isothermal \citep{Sanderson06} within
$\sim100$ kpc and show little metallicity gradient \citep{DeGrandi01}.
CC clusters generally have central cooling times of $0.1-1.0$ Gyr
while the cooling times in NCC clusters are generally somewhat higher
at $\sim1$ Gyr \citep{Sanderson06}.  The present work investigates the
expected thermal properties of both CC and NCC clusters.

The aim of the present study is not to find one or more heating
mechanisms that can offset radiative cooling but rather to understand
the general conditions that lead to thermal balance in the ICM for a
variety of possible mechanisms.  This requires both a heating
mechanism (or mechanisms) that can energetically offset radiation, and
also maintain long term thermal balance.  

The rest of this paper proceeds as follows.  In $\S$\ref{sec:methods}
the methods are discussed, including the implementation of cooling,
supernovae heating, conduction, DF heating, and relativistic pressure,
the initial equilibria, and the numerical setup.  $\S$\ref{sec:nexp}
contains the results of a series of numerical experiments where the
thermal balance of the ICM is investigated.  A discussion and summary
of these results can be found in $\S\S$\ref{sec:disc} and
\ref{sec:sum}, respectively.  Throughout we assume $h=0.7$ where $h$
is the Hubble parameter in units of $100$ km s$^{-1}$ Mpc$^{-1}$.

\section{Methods \& Physical Processes}
\label{sec:methods}

This section reviews the relevant fluid equations, cooling function,
and heating mechanisms that will be explored in depth in the following
sections.  This section also discusses our implementation of a
relativistic fluid component, initial equilibria, and the numerical
setup.

\subsection{General Equations}
\label{sec:gen}

The hydrodynamical equations are:
\be
\frac{d \rho_g}{d t} = - \rho_g {\bf \nabla \cdot \bf v},
\label{e:hydro1}
\ee
\be
\frac{d{\bf v}}{d t} = -\frac{1}{\rho_g}{\bf \nabla}P_{tot} + \, {\bf g},
\label{e:hydro2}
\ee
\be
\frac{de_g}{d t} = - (e_g+P_g) {\bf \nabla \cdot v} + \Gamma - \Lambda,
\label{e:hydro3}
\ee
\noindent
where $\rho_g$, $e_g$, $P_g$, $P_{\rm tot}$ and ${\bf v}$ are the gas
density, internal energy density, gas pressure, total pressure, and
velocity, ${\bf g}$ is the total gravitational acceleration, and
$\Gamma$ and $\Lambda$ are the heating and cooling functions per unit
volume.  We have explicitly distinguished between the total and gas
pressures because below we will allow for the addition of a
relativistic fluid that can provide additional pressure support. The
equation of state for the gas is:
\noindent
\be
e_g = \frac{1}{\gamma_g-1} P_g,
\ee
where $\gamma_g=5/3$ is the ratio of specific heats.

The gravitational acceleration is the combination of a (passive) dark
matter halo, a central cD galaxy, and the self-gravity of the gas:
\be
g(r) = g(r)_{\rm{DM}} + g(r)_{\rm{cD}} - \frac{G\,M_{\rm{gas}}(<r)}{r^2},
\ee
where $g_{\rm{DM}}$ is
\be
g(r)_{\rm{DM}} = -\frac{2GM_0}{r_s^2}\bigg[\frac{\rm{ln}(1+x)}{x^2} - \frac{1}{x(1+x)}\bigg],
\ee
and is derived from the NFW \citep{NFW97} density profile:
\be
\rho_{\rm{DM}} = \frac{M_0/2\pi}{r(r+r_s)^2},
\ee
\noindent
where $G$ is Newton's constant, $x\equiv r/r_s$, $M_0$ is the
normalization and $r_s$ is the scale radius, i.e. the radius at which
the density profile scales as $r^{-2}$.  For reference, the mass
within $2r_s$ is equal to $0.86 M_0$ for the above density profile.
We fix $r_s=460$ kpc for comparison to \citet{Kim05}.  This value for
$r_s$ is bracketed by the observational range \citep{Vikhlinin06a}.
$M_0$ is allowed to vary as discussed in $\S$\ref{s:ie}.

The central cD galaxy mass density profile is taken to be a King
profile:
\noindent
\be
\rho_{\rm{cD}}(r) = \frac{\rho_{\rm{cD},0}}{[1+(r/r_{\rm{cD}})^2]^{3/2}},
\label{e:cd}
\ee
where
\be
\rho_{\rm{cD},0} = \frac{9 \sigma_{\rm{cD}}^2}{4\pi G r_{\rm{cD}}^2}.
\ee
\noindent
In the above equations, $\sigma_{\rm{cD}}$ is the central 1D velocity
dispersion of the central galaxy and $r_{\rm{cD}}$ is the core radius
for the King profile; these parameters are taken to be 200 $\kms$ and
2.83 kpc, respectively, from a fit to the cD galaxy NGC 6166
\citep{Kelson02} which is representative of cD stellar density profiles.
Note that the King profile is for all practical purposes quite similar
to the more conventional de Vaucouleurs profile, but is more
analytically tractable.  The gravitational acceleration associated with
this density distribution is
\noindent
\be
g_{\rm{cD}}(r) = -\frac{9\sigma_{\rm{cD}}^2}{r_{\rm{cD}}} \bigg(\frac{1+r'(1+r'^2)^{-0.5}}{r'(r'+\sqrt{1+r'^2})} - \frac{{\rm ln}(r'+\sqrt{1+r'^2})}{r'^2} \bigg),
\ee
where $r'\equiv r/r_{\rm{cD}}$.

Mass that flows through the inner boundary (1 kpc; see below), and is
hence no longer followed directly in the simulation, is added to the
cD galaxy.  Our results are unchanged if the mass is added to a
central black hole instead.

The gas is assumed to be ideal:
\be
P = \frac{\rho_g k_B T}{\mu m_\mu} = \frac{\mu_e}{\mu}n_e k_B T,
\ee
\noindent
where $\mu$ is the mean molecular weight, $m_\mu$ is the atomic mass
unit, $k_B$ is Boltzmann's constant, $n_e$ is the electron number
density and $T$ is the gas temperature.  Throughout, the gas
metallicity is taken to be $(1/3) \,Z_\Sun$ (except in $\S$\ref{s:ie}
where we discuss the sensitivity of our results to different
metallicities).

\subsection{Heating \& Cooling}
\label{sec:phy}

This section describes the various heating and cooling mechanisms that
are potentially relevant for the thermodynamics of the ICM.  Each
mechanism includes an adjustable free parameter that is meant to
encapsulate both our ignorance regarding the applicability of the
mechanism to real clusters and the uncertain values of the particular
parameters relevant for each mechanism; a summary of these parameters
is provided in Table \ref{t:models}.

\subsubsection{Radiation}

\begin{figure}[t]
\plotone{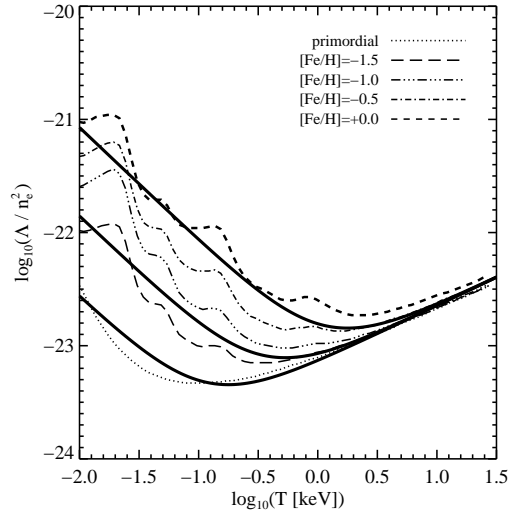}
\vspace{0.5cm}
\caption{Cooling function used in the simulations (\emph{solid lines})
  compared to the more accurate cooling functions of
  \citet{Sutherland93}.  The solid lines are, from bottom to top, for
  primordial metallicity, $[Fe/H]=-1.5$ and $[Fe/H]=0.0$.  Note that
  even for low metallicity the cooling function deviates from pure
  free-free cooling (where $\Lambda \propto T^{1/2}$) at
  $T\gtrsim10^7$ K due to recombination cooling.}
\vspace{0.5cm}
\label{fig:coolfn}
\end{figure}

A simplified cooling function is adopted that broadly captures the
metallicity dependent features in the detailed cooling function of
\citet{Sutherland93} via the following function:
\noindent
\be
\label{eqn:rad}
\Lambda = 2.1 \times 10^{-27} \,n_e^2\, T^{1/2}\bigg[1+\bigg(\frac{1.3\times 10^7\, \zeta(Z)}{T}\bigg)^{1.5}\bigg],
\ee
\noindent
in units of $\rm{erg} \,\rm{s}^{-1} \rm{cm}^{-3}$.  The variable
$\zeta$ is a simple function of the gas metallicity, $Z$.  This
metallicity-dependent cooling function is shown in Figure
\ref{fig:coolfn} (solid lines) along with the detailed cooling
functions from \citet{Sutherland93}.  The cooling function is
truncated at $T=0.01$ keV because our analytic approximation becomes
inaccurate at lower temperature, and because more complex physical
phenomena not included herein (such as star formation) become relevant
at such temperatures.  Most of our results focus on a single
metallicity of $Z=\frac{1}{3} \,Z_\Sun$; in $\S$\ref{s:ie} we briefly
discuss how the results change when different metallicities are adopted.

\subsubsection{Supernovae}

The stellar populations of central galaxies are extremely old, with
typical formation epochs at $z>3$ \citep{Thomas05}; there are thus no
type II supernovae events in these old systems.  However, type Ia
supernovae events are common, even for the old stellar populations
comprising cD galaxies.  At low redshift, roughly one type Ia
supernovae event occurs every 100 years per $10^{12} M_\Sun$ of old
stars \citep{Mannucci05}.  Since each supernova releases $\sim10^{51}$
ergs, which we assume is transfered entirely to the ICM, this
corresponds to a time-averaged energy injection rate of $10^{49}$ erg
yr$^{-1}$ per galaxy of mass $10^{12} M_\Sun$.  The type Ia rate for a
co-evolving stellar population may be a decreasing function of time,
indicating that the rate at earlier epochs would be higher than the
value we adopt here \citep{Greggio05, Mannucci06}.  For our purposes
we are primarily interested in type Ia rates at $>1$ Gyr after the
formation epoch of the stars, precisely where the time-dependence of
the rates are least certain \citep{Mannucci06}.  However, even a
modest increase of the rate with redshift would not qualitatively
change our conclusions.  We therefore do not include any
time-dependence in the type Ia rate in what follows.

Since there are hundreds of galaxies within clusters,
supernovae-related energy injection should be treated throughout the
cluster.  However, since radiation is capable of substantially cooling
only the inner cluster region (because cooling scales as $\rho^2$ at
fixed temperature), and since the thermal reservoir in the outer
regions is large, we only distribute supernovae within the central
galaxy for simplicity.  The supernovae energy injection rate is
distributed with the same space density as the cD galaxy (cf. Equation
\ref{e:cd}).

Supernovae energy injection is non-negligible in the central regions
at the beginning of our simulations.  For a initially isothermal
cluster with central electron density $0.020$ cm$^{-3}$ and $T_i=6$
keV, the volume-averaged supernovae heating initially dominates
cooling within only 2 kpc, while for a central density of $0.005$
cm$^{-3}$ supernovae heating dominates within $\sim10$ kpc.  However,
since the injection rate is independent of the properties of the ICM,
it has little impact on the long-term thermal balance of the ICM.  In
particular, while for very low density intracluster media supernovae
heating is comparable to radiative cooling on small scales, it is
completely ineffective at larger scales (i.e. several tens of kpc)
because there are so few stars, and hence so few supernovae, at these
larger scales.

We have run many simulations for a variety of initial conditions and
additional heating sources (discussed below) with supernovae feedback
and indeed find that while at the central regions ($\sim1-10$ kpc) it
can be important, it has little effect on the long-term thermal
balance of the ICM.  These conclusions are qualitatively similar to
those of \citet{Kravtsov00} who used the observed metallicity of the
ICM to demonstrate that supernovae heating is insufficient to offset
radiative losses.  In order to simplify the discussion in the
following sections, we neglect this energy injection process for the
remainder of this work.

\subsubsection{Conduction}

The heat flux due to electron conduction may be described by:
\be
\Gamma_{\rm cond} = {\bf \nabla \cdot}(\kappa {\bf \nabla}T),
\ee
\noindent
where $\kappa$ is the conductivity.  For a fully ionized plasma, the
conductivity is:
\noindent
\be
\label{eqn:cond}
\kappa = f\,\kappa_{Sp} = f\, \frac{1.84\times 10^{-5}\, T^{5/2}}{\rm{ln}\Lambda_C} \,\,\,\, \rm{erg}\,\rm{s}^{-1}\, \rm{K}^{-1} \, \rm{cm}^{-1},
\ee
\noindent
where $\kappa_{Sp}$ is the classical \citet{Spitzer62} conductivity,
$\rm{ln}\Lambda_C\sim37$ is the Coulomb logarithm, and $f$ is a free
parameter describing the suppression of conductivity relative to the
full Spitzer value.  While it is difficult to constrain the globally
averaged value of $f$ in observed clusters, the strength of observed
temperature gradients in several clusters suggests that $f\ll 1$
\citep{Markevitch03, Xiang07}.

There are many theoretical reasons to expect that $f<1$.  The primary
uncertainty on the importance of conduction is the strength and order
of magnetic fields in the ICM.  Magnetic fields act to suppress
conduction across magnetic field lines (due to the small gyro-radius
of electrons) while permitting conduction along the field lines.
Observations suggest that the magnetic fields in clusters are tangled
with length scales $\sim1-10$ kpc \citep[e.g.][]{Taylor02}.  Tangled
magnetic fields or other magnetic phenomena tend to suppress the
thermal conductivity to a value roughly $10-30$\% of the fully Spitzer
value \citep[e.g.][]{Malyshkin01, Narayan01, Chandran04}, although
turbulence may boost the effective conductivity \citep{Cho04}.  The
amount of suppression depends in detail on the poorly constrained
properties of the cluster magnetic field, and should thus be
considered rather uncertain.  Finally, it should be kept in mind that
three-dimensional simulations of conduction in the presence of
magnetic fields display new instabilities \citep[e.g.][]{Parrish07a}
that cannot be captured in our one-dimensional treatment where the
conduction is by necessity isotropic.  These considerations lead us to
adopt $0.0<f<0.5$ as a plausible range for suppression of
isotropic conduction (see Table \ref{t:models}).

\subsubsection{DF Heating}\label{s:df}

A wealth of energy is stored in the orbital motions of galaxies within
clusters.  DF provides a way to transfer that energy to the background
matter, including both the dark matter and the ICM, and can be a
potential mechanism that balances the cooling flow, as first pointed
out by \citet{Miller86}.  Recently, DF heating has experienced a
resurgence in popularity \citep{ElZant04b, Kim05, Kim07} thanks in
part to detailed calculations of the efficiency of DF in a collisional
medium \citep{Ostriker99}, which have been verified by controlled
numerical experiments \citep{Sanchez01, Kim07a}. These calculations
showed that DF heating is stronger in collisional media (such as the
ICM) when galaxies are moving slightly supersonically, as appears to
be the case in clusters \citep{Faltenbacher05}, where the average Mach
number of galaxies is $\sim 1.3$.  DF will also heat the background
dark matter halo, thereby producing a core-like inner density profile,
as opposed to the cusp implied by the NFW distribution
\citep{ElZant04a}, although this appears to depend on the detailed
properties of the accreted objects \citep{Boylan-Kolchin07}.

A simple calculation suggesting that DF heating may be important in
real clusters will be presented first, followed by a more detailed
discussion of our implementation of DF heating.

\bigskip

$X$-ray luminosities of clusters vary from $\sim10^{43}$ to
$\sim10^{45}$ ergs s$^{-1}$, with of order $10$\% emitted within the
rapidly cooling inner region \citep{Fabian94}.  Assuming an average
$X$-ray luminosity of $10^{44}$ ergs s$^{-1}$, this implies a loss of
energy from $X$-ray radiation over a Hubble time of
$\sim5\times10^{60}$ ergs within the cooling inner region
\citep[e.g.][]{Fabian94}.  The energy liberated during the build-up of
the massive central cD galaxy is a plausible source of energy that may
balance these radiative losses.  The following is a simple calculation
demonstrating that the binding energy of the cD absorbed by the ICM is
of the same order as the energy radiated away within the cooling
region.

The energy liberated during the build-up of the cD galaxy is
\be
\Delta E = \int \phi \,{\rm d}m_\ast = 4\pi \, \int \phi \rho_\ast\,r^2 {\rm d}r,
\ee
\noindent
where $\Delta E$ is the binding energy, $\phi$ is the total
gravitational potential felt by the cD, $m_\ast$ is a unit of stellar
mass, and $\rho_\ast$ is the stellar density profile.  We assume that
$\phi$ is static for simplicity, though of course the dark matter halo
is being built-up at the same time as the cD.

The cD galaxy projected density profile is approximated as a Sersic
profile \citep{Sersic68}, in agreement with a variety of observations.
The projected Sersic profile is not analytically invertible, but
sufficiently accurate fitting functions exist in the literature.  The
de-projected stellar mass density can be approximated by
\citep{LimaNeto99}:
\noindent
\be 
\rho_\ast(r) = A \, s^{-\alpha}\,{\rm exp}\big(-b_ns^{1/n}\big),
\ee
\noindent
where $s \equiv r/R_e$, $b_n \approx 2n-0.327$, and $\alpha\approx
1-1.188/(2n) + 0.22/(4n^2)$.  $A$ is an integration constant chosen
such that the integral over $\rho_\ast$ equals the total stellar mass
$M_\ast$.

For simplicity the total gravitational potential within the inner
regions is assumed to be isothermal:
\noindent
\be
\phi(r) = {\rm V}_c^2 \, {\rm ln}(r/r_{\rm cool}),
\ee
where ${\rm V}_c$ is the circular velocity.  We are only interested in
energy liberated within the cooling region ($r<r_{\rm cool}$), and so
set the potential to be zero at $r=r_{\rm cool}$.  Define
$\gamma\equiv r_{\rm{cool}} / R_e$.

Combining these results yields:
\be
\Delta E =  V_c^2\, M_\ast \, \, \frac{\int_s^0 {\rm ln}\big(\frac{s'}{\gamma}\big) \,s'^{2-\alpha} \,{\rm exp}\big[-b_n s'^{1/n}\big]{\rm d}s'}{\int_0^s \,s'^{2-\alpha} \,{\rm exp}\big[-b_n s'^{1/n}\big]{\rm d}s'}.
\ee
\noindent
The two integrals represent an average of the logarithmic run of the
potential over the stellar density profile.  The integrals are
functions only of the Sersic index $n$, the ratio between the cooling
radius and the effective radius $\gamma$, and the upper limit of
integration $s$ (taken to be the same in both integrals, which is
reasonable but not necessary).  For $1<s<3$ and $4<n<10$ the ratio of
these two integrals varies from $0.4$ to $1.8$, for $\gamma=1$.  The
integral scales roughly as $\sqrt{\gamma}$.

For a typical cD galaxy $M_\ast\sim 10^{12} \,M_\sun$ and for a typical
cluster ${\rm V}_c\sim 700\kms$ \citep{Gonzalez07} implying that $\Delta
E\sim 10^{61}$ ergs.  This energy can be transferred into both the
background ICM and dark matter halo.  If the gas were collisionless,
then the fraction of energy shared between the gas and the dark matter
would be simply proportional to their relative mass fractions
($\sim1/6$ and $\sim5/6$ for the gas and dark matter respectively).
However, the collisional nature of the gas makes DF more efficient at
transonic speeds \citep{Ostriker99}; at Mach numbers near unity the
efficiency of DF heating on the gas is roughly twice that of the
collisionless dark matter.  This implies that $\sim30$\% of the energy
released in transferred to the ICM and the rest to the dark matter.
The cD binding energy thus absorbed by the ICM is comparable to the
energy radiated away within the cooling inner region over a Hubble
time.

There were many simplifications made in this calculation.  One
complication is that observations indicate that a large fraction of
massive ($M_\ast>10^{11.5} M_\Sun$) galaxies were already in place by
$z\sim1$ \citep[e.g.][]{Bundy05, Borch06, Fontana06, Brown07}, and
hence the energy liberated from the build-up of cDs may not provide
much energy to the ICM at $z<1$.  However, the outer envelope of the
cD may be growing substantially at $z<1$ from the shredding of
satellite galaxies \citep[see e.g.][for a discussion]{Ostriker77,
  Conroy07b, Purcell07}, suggesting that energy released from the
build-up of the cD is available at late times.  Moreover, the strong
small-scale clustering of luminous red galaxies indicates that there
are many massive galaxies orbiting near the cluster center, plausibly
transferring their orbital energy to the ICM and background halo via
DF \citep{Masjedi06}.

These calculations suggest that DF heating may be an important heating
mechanism.  While the simulations below address DF heating in more
detail, detailed high-resolution three-dimensional simulations are
required to fully address the importance of DF heating in the ICM of
clusters \citep[see e.g.][]{Kim07}.  Moreover, detailed resolution
studies are required to address whether or not current cosmological
hydrodynamic simulations are adequately resolving DF \citep[see
discussion in][]{Faltenbacher05, Naab07}.

\bigskip

Our implementation of DF heating closely follows that of
\citet{Kim05}; the reader is referred to that work for more details.
Note that we do not track orbits explicitly nor do we attempt to
resolve the actual DF wake.  We take a more approximate,
parameterized approach to the spherically-averaged heating rate due
to DF.

The heat flux due to DF can be described as:
\noindent
\be
\label{eqn:df}
\Gamma_{DF} = n_{\rm{gal}} \langle -{\bf F_{\rm{DF}} \cdot v} \rangle =
d \frac{4\pi \rho_g G^2  \overline{M_{\rm{gal}}^2} \langle I/\mathcal{M} \rangle}{c_s} \, n_{\rm{gal}}(r),
\ee
\noindent
where $\mathcal{M}\equiv {\bf v}/c_s$ is the Mach number,
$n_{\rm{gal}}$ is the number density of galaxies,
$\overline{M_{\rm{gal}}^2}$ is the average squared total mass of the
galaxies, and $c_s$ is the isothermal sound speed.  In what follows we
will argue for reasonable values for each of the parameters in
Equation \ref{eqn:df}, and will then incorporate the uncertainties and
possible ranges in all of these parameters into the single free
parameter $d$.  The angular brackets indicate an average over the
velocity distribution function.

Note that $M_{\rm gal}$ includes both the stellar and dark matter of
the satellite galaxies, and that since DF is proportional to the
average of the galaxy mass {\it squared}, more massive galaxies are
given greater weight.  The quantity $\overline{M_{\rm{gal}}^2}$ is
estimated as follows.  We assume that the dark matter-to-stellar mass
ratio is constant for satellites, and estimate it by taking the total
stellar mass of satellites in clusters to be 1\% of the total cluster
mass \citep{Gonzalez07} and the total amount of dark matter mass in
substructures to be 10\% \citep{Gao04b}, where the substructures are
assumed to be the likely locations of satellite galaxies
\citep{Conroy06a}.  These numbers imply an average dark
matter-to-stellar mass ratio of 10 for satellite galaxies.  If the
total mass is $M_{\rm gal}$, then $\overline{M_{\rm{gal}}^2} =
\overline{M_{\rm{star}}^2} + \overline{M_{\rm{dm}}^2} =
101\,\overline{M_{\rm{star}}^2}$.  The stellar mass function provides
an estimate of $\overline{M_{\rm{star}}^2}$ and thus of
$\overline{M_{\rm{gal}}^2}$.  We adopt the global mass function of
\citet{Bell03} where $\alpha=1.1$ and ${\rm log}(M^\ast)=10.9\,M_\Sun$
are the best-fit Schechter parameters.  The Schechter parameters for
the luminosity function within clusters does not differ strongly from
the global value \citep{Hansen07}, and so we adopt the global values
for this calculation.  Integrating the stellar mass function to
$M_{\rm star}=10^8\,M_\Sun$ leads us to adopt
$\overline{M_{\rm{gal}}^2} = (10^{11}\,M_\Sun)^2$ as the fiducial
value.  The uncertainty on this quantity is explored via the tunable
parameter $d$ discussed above.

By definition we have:
\noindent
\be
\langle I/\mathcal{M} \rangle = \frac{\int I/\mathcal{M} f({\bf v}) d{\bf v}}{\int f({\bf v}) d{\bf v}},
\ee
where $f$ is assumed to be Maxwellian \citep{Faltenbacher05}:
\be
f({\bf v}) = \frac{4\pi N_{\rm{gal}}}{(2\pi\sigma_r^2)^{3/2}}\,{\bf v}^2\, e^{-{\bf v}^2/(2\sigma_r^2)},
\ee
\noindent
where $\sigma_r$ is the radial velocity dispersion of galaxies
(assumed for simplicity to be independent of radius), and
$N_{\rm{gal}}$ is the total number of galaxies within the cluster.  In
the following we take $\sigma_r =1000$ km s$^{-1}$ and
$N_{\rm{gal}}=500$.  These parameters are fixed throughout; any
variation of these physical quantities is incorporated into the $d$
parameter.  Note that DF heating depends on the state variables only
through $c_s$, $I/\mathcal{M}$, and $\rho_g$, and that
$\Gamma_{DF}\propto \rho_g$ at fixed temperature.  $I$ is the
efficiency factor for DF in the collisional case \citep{Ostriker99}:
\noindent
\begin{displaymath}
I \equiv \left\{ \begin{array}{ll}
 \frac{1}{2}\,\rm{ln}(1-\mathcal{M}^{-2}) + \rm{ln}(vt/r_{\rm{min}}) & \mathcal{M} >1 \\
\frac{1}{2}\,\rm{ln}\bigg(\frac{1+\mathcal{M}}{1-\mathcal{M}}\bigg) -\mathcal{M} & \mathcal{M} <1
\end{array} \right.
\end{displaymath}
\noindent
As is common practice, we set $vt=r_{\rm{max}}$, where $r_{\rm max}$
and $r_{\rm min}$ are in this problem taken to denote the size of the
cluster and the satellite galaxies, respectively.  The factor ${\rm
  ln}(vt/r_{\rm{min}})$ thus plausibly varies from $\sim4-10$.  In
what follows we set this factor to 6 and note that our results are
insensitive to this particular value.  Note that DF heating is more
efficient at subsonic speeds than one might naively expect because the
relevant efficiency is $\langle I/\mathcal{M} \rangle$, which, though
maximal at $\mathcal{M}=1$, decreases by only $50$\% at
$\mathcal{M}\approx0.5$ and decreases only weakly at $\mathcal{M}> 1$.
This implies that the feedback provided by the Mach number (in the
sense that colder/hotter systems will have more/less efficient DF
heating) is weaker than one might have expected.

The number density of galaxies, $n_{\rm{gal}}(r)$, is taken to
be a modified King profile with the parameters adopted from \citet{Girardi98}:
\noindent
\be
n_{\rm{gal}}(r) = n_{\rm{gal}}(0)[1+(r/r_c)^2]^{-1.2},
\ee
\noindent
where the core radius $r_c=50$ kpc.  The central density is set by
requiring that $n_{\rm{gal}}(r)$ integrate to the total number of
galaxies, $N_{\rm{gal}}$.  Note that $n_{\rm{gal}}(r)$ does not
include the central cD galaxy.  While most cD galaxies are near the
center of the halo as defined by $X$-ray imaging \citep{Lin04b}, the
cD may be oscillating about the gravitational center with an amplitude
of several kpc \citep{VDB05b}.  Such oscillations may provide
additional heating in the central regions, but we do not include them
herein.

\begin{deluxetable}{lll}
\tablecaption{Summary of Free Parameters}
\tablehead{ \colhead{} & \colhead{} &\colhead{Plausible} \\ 
 \colhead{Parameter} & \colhead{Comment} &\colhead{Range}}
\startdata\\
$Z/Z_\Sun$ & Metallicity of the gas (Eqn. \ref{eqn:rad}) & $0.1-0.6$\\
$f$ & Fraction of Spitzer conductivity  (Eqn. \ref{eqn:cond}) & $0.0-0.5$ \\
$d$ & Normalization of DF heating  (Eqn. \ref{eqn:df}) & $0.1-10.0$\\
$\alpha$ & Initial fraction of relativistic pressure  (Eqn. \ref{eqn:nth}) & $0.0-0.3$\\
\enddata
\label{t:models}
\end{deluxetable}
\vspace{0.5cm}

\subsection{Relativistic Pressure}
\label{s:prel}

Relativistic pressure in the form of cosmic rays could be an important
dynamical component of the ICM.  It has long been known that
astrophysical shocks efficiently accelerate cosmic rays
\citep{Blandford78}.  Numerical simulations suggest that cosmic rays
accelerated in shocks could account for $\sim10$\% of the total
cluster pressure \citep{Miniati01, Ryu03}.  Cosmic rays may also be
generated in buoyantly rising bubbles generated by AGN activity
\citep[e.g.][]{Ensslin03}.  Observations suggest that a relativistic
component may constitute several tens of percent of the total energy
density of clusters \citep{Pfrommer04, Sanders05, Dunn06, Sanders07,
  Werner07, Nakar07}.  In addition, it has been shown that a
relativistic pressure component can suppress thermal instabilities
\citep[][see also the appendix herein]{Cen05}.

There have been several recent attempts to study the effects of
cosmic-rays in the ICM numerically \citep{Pfrommer07a, Pfrommer07b,
  Pfrommer07c, Guo07}.  The model of \citet{Guo07}, which includes
both conduction and cosmic rays injected into the ICM from AGN-induced
bubbles, is able to reproduce the temperature and density profiles of
observed clusters.  Perhaps the most attractive feature of their model
is that their results do not require fine tuning of the various
adjustable parameters, including the amount of thermal conductivity.
It would be interesting to know whether and to what extent their
results rely on the relativistic pressure provided by the CRs, or
whether it is due primarily to the energetics associated with the
bubbles.  In the present work we test the former hypothesis.

In the following we take a simplified approach when exploring the
effects of a relativistic component.  The relativistic pressure
($P_r$) is assumed to be a fixed fraction of the total pressure
initially:
\noindent
\be
\label{eqn:nth}
P_r = \alpha P_{tot},
\ee
\noindent
and we further assume that the relativistic component is perfectly
dynamically coupled to the gas (i.e. there is no motion or diffusion
of one component relative to another), and evolves adiabatically
(i.e. $P_r\propto \rho_g^{\gamma_r}$).  These requirements lead to a
fourth fluid equation:
\noindent
\be
\frac{\partial P_r}{\partial t} = - {\bf \nabla \cdot}({\bf v}P_r) + (1-\gamma_r)\,P_r{\bf \nabla \cdot} {\bf v},
\ee
\noindent
which is simply a statement of energy conservation. In the above
equation $\gamma_r=4/3$ and we have made use of the following equation of state:
\be
e_r = \frac{1}{\gamma_r-1} P_r,
\ee
\noindent
where $e_r$ is the energy density associated with the relativistic
fluid.  Note that relativistic pressure also enters into the
hydrodynamical equations (Equations \ref{e:hydro1}---\ref{e:hydro3})
by contributing to the total pressure.

These simple assumptions are motivated by CR creation and evolution in
real clusters.  It may be the case that CRs are created predominately
in the cluster center via AGN activity, or it may be that the
accretion shock at the cluster outskirts are the predominate source of
CRs.  CRs are capable of diffusing out of the cluster, but CR loss via
diffusion may be approximately balanced by the creation of new CRs
throughout the cluster.  Our simplified treatment is meant to
demonstrate the potential importance of CRs generally (insofar as they
provide relativistic pressure support); it will be the task of more
sophisticated modeling efforts \citep[see e.g.][]{Pfrommer07a,
  Pfrommer07b, Pfrommer07c, Guo07} and, ultimately, observations to
refine our knowledge of CR production and evolution.  Note that the
energetics associated with CR production (whether in e.g. shocks or
AGN-related bubbles) is not explored herein.

\subsection{Initial Equilibria}
\label{s:ie}

Every simulated cluster is set up initially in hydrostatic
equilibrium.  However, in general the clusters are {\it not} set up in
thermal balance.

Observations indicate that there are two classes of intracluster
media, the cooling-core (CC) and non-cooling-core (NCC) clusters
\citep{Sanderson06}.  Clusters in the former class show a rapidly
declining temperature profile with decreasing radius at radii less
than $\sim 100$ kpc, while the latter are approximately isothermal
within the same physical region.  Furthermore, CC clusters have factors
of $\sim2-3$ metallicity gradients within $\sim 100$ kpc while NCC
clusters do not \citep{DeGrandi01}.  For our analysis we assume that
neither NCC nor CC clusters have an appreciable metallicity gradient,
and that the metallicity is $(1/3)Z_\Sun$.  We have selected several
cluster runs at random and re-simulated them with metallicity varying
from primordial to solar composition and find no qualitative change in
our results.

\begin{figure}[t]
\plotone{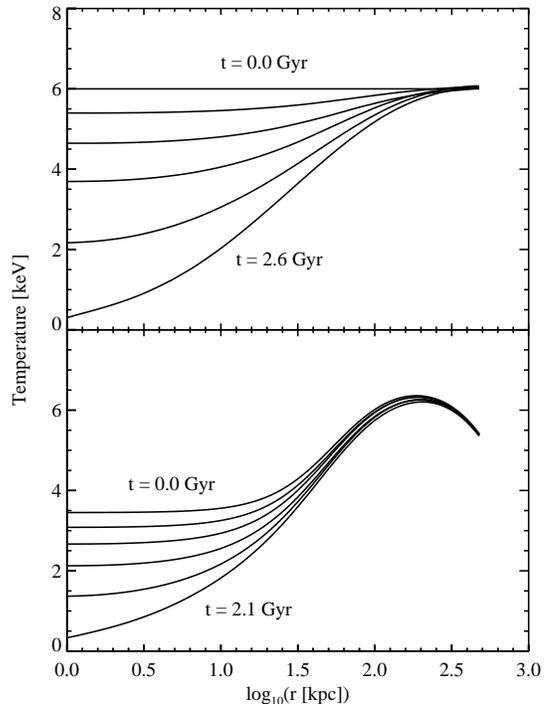}
\caption{Evolution of the temperature profile of clusters with
  radiative cooling and no heating for $n_{e,0}=0.02$ cm$^{-3}$.  As
  expected, a runaway cooling flow develops and the core collapses
  within a cooling time.  \emph{Top Panel:} An initially NCC cluster.
  \emph{Bottom Panel:} An initially CC cluster.  The lines are spaced
  equally in time.}
\vspace{0.5cm}
\label{fig:rad_run}
\end{figure}

Both of these classes are discussed in the sections that follow.  NCC
clusters as approximated as initially isothermal while the CC clusters
are assumed to have an initial temperature profile that is described
by \citep{Vikhlinin06a}:
\noindent
\be T(r)/T_{\rm CC} =  \frac{1.15\,(x/0.045)^2 + 0.5}{(x/0.045)^2+1}  \frac{1}{1+(x/0.75)^2},
\ee
\noindent
where $x\equiv r/r_{500}$.  This temperature profile peaks at $T_{\rm
  CC}$ where $r\sim0.1 r_{500}$ and drops by a factor of $\sim2-3$
toward the center.  Figure \ref{fig:rad_run} (\emph{bottom panel})
provides an example of a CC cluster temperature profile.

In each run the cumulative gas fraction at $500$ kpc is fixed at
$f_g=0.1$, in agreement with recent observations \citep{Sanderson03,
  Vikhlinin06a}.  The gas fraction is less than the universal baryon
fraction of $f_b=0.16$ for several reasons \citep[see e.g.][for a
discussion]{Kravtsov05}.  The conversion of gas into stars results in
a lower gas fraction by several tens of percent \citep{Fukugita98,
  Lin03}.  Both adiabatic effects and heating processes further
decrease the gas fraction within $\sim500$ kpc \citep{Bode07}.  Fixing
the gas fraction implies that simulated clusters with larger central
electron densities are embedded in larger dark matter halos, as
observed \citep{Vikhlinin06a}.  

With the above simplifications, each simulation is fully specified by
the initial central electron density and the temperature profile.
These two parameters, along with the assumption of hydrostatic
equilibrium, then determines the full density, temperature, and
pressure profiles, and the adopted gas mass fraction then determines
the mass of the static dark matter halo.  The contribution of the cD
galaxy to the gravitational potential is fixed initially and increases
in proportion to the amount of mass that flows through the inner
boundary condition (see $\S$\ref{sec:gen}).  In practice we
iteratively solve for $M_0$ such that hydrostatic balance is achieved.
For the central densities explored below, the NFW normalization $M_0$
ranges from $\sim(2-7)\times10^{14} M_\Sun$.  Allowing $M_0$ to vary
by fixing $f_g$ as opposed to allowing $f_g$ to vary by fixing $M_0$
has a negligible effect on our conclusions.

Observationally the temperature of clusters varies from $\sim1-10$
keV, and is strongly correlated with the total cluster mass
\citep[e.g.][]{Vikhlinin06a}.  The data also cover a wide range in
central electron densities, roughly $10^{-2.5}<n_{e,0}<10^{-0.5}$
cm$^{-3}$.  We pay more attention to how cooling and heating depends
on density because the cooling and heating functions are more
sensitive to density than temperature.  Below we include ``observed''
central electron densities for comparison with our initial conditions,
which are estimated from two sources.  \citet{Zakamska03} provide
central densities from hydrostatic fits to observed temperature and
density profiles of 10 clusters.  \citet{Vikhlinin06a} provide
detailed parametric fits to the electron density profiles of 13
clusters (two clusters are in both samples and are only counted once
herein).  For the latter sample we quote as the central density the
density at 10 kpc because in many cases the density profile at smaller
scales is enhanced due to cooling.  Since we are interested in
``initial'' central electron densities, we essentially mask out this
inner cooling region when comparing to the initial electron densities
in our simulations.  In the simulations the initial central electron
densities are approximately constant within the inner 10 kpc; we thus
consider this simplification appropriate.

\subsection{Numerical Setup}

We utilize the Lax-Wendroff method \citep{Press92} to integrate the
fluid equations in their one-dimensional form.  The spatial grid is
logarithmic and has $N=400$ elements with range $1<r<500$ kpc.  We
have doubled the number of grid elements for several simulations and
extended the spatial grid to 1000 kpc and find no change in the
results.  The time-step is determined by the Courant condition:
\noindent
\be
\Delta t_{CFL} = 0.5\, min\bigg( \frac{\Delta r_i}{|{\bf v}_i+c_{s,i}|} \bigg),
\ee
where $i=[0,N]$ and $c_{s,i}$ is the local isothermal sound speed.

Conduction is implemented with a fully implicit algorithm
\citep{Press92}.  The pressure, density, and temperature in the ghost
cells are linearly extrapolated from the active zones, both at the
inner and outer boundaries.  At the inner boundary, the velocity is
also a linear extrapolation from the active zones unless the velocity
in the ghost cell is positive (indicating outflow) in which case the
velocity is set to zero.  The velocity in the ghost cell at the outer
boundary is always zero.

\section{Numerical Experiments}
\label{sec:nexp}

This section presents the results of a series of numerical
experiments.  The various physical processes discussed in
$\S$\ref{sec:phy} are discussed in a variety of combinations in order
to elucidate their various effects.  For each process we discuss the
sensitivity of the final state to the free parameters summarized in
Table \ref{t:models}.

\subsection{Radiation Only}

\begin{figure}[!t]
\plotone{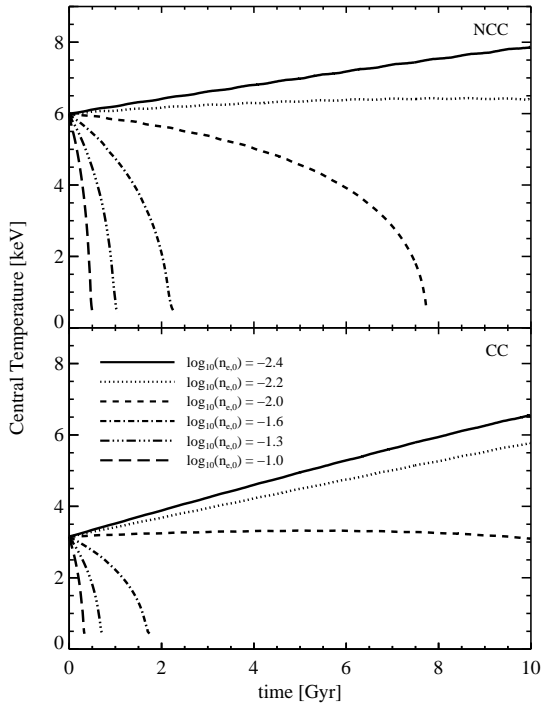}
\vspace{0.5cm}
\caption{Evolution of the central temperature for simulations that
  include both radiation and DF heating.  Each panel shows the
  evolution of clusters with a variety of initial central densities
  (labeled in the figure in units of cm$^{-3}$).  The normalization of
  DF heating here is fixed at $d=1.0$ (cf. Equation \ref{eqn:df}).
  \emph{Top Panel:} Initially NCC clusters.  \emph{Bottom Panel:}
  Initially CC clusters.  The evolution of CC clusters is
  qualitatively similar to NCC clusters in the sense that thermal
  balance can only be maintained for a narrow range of central
  electron densities.}
\vspace{0.5cm}
\label{fig:rad_df}
\end{figure}

\begin{figure*}
\plottwo{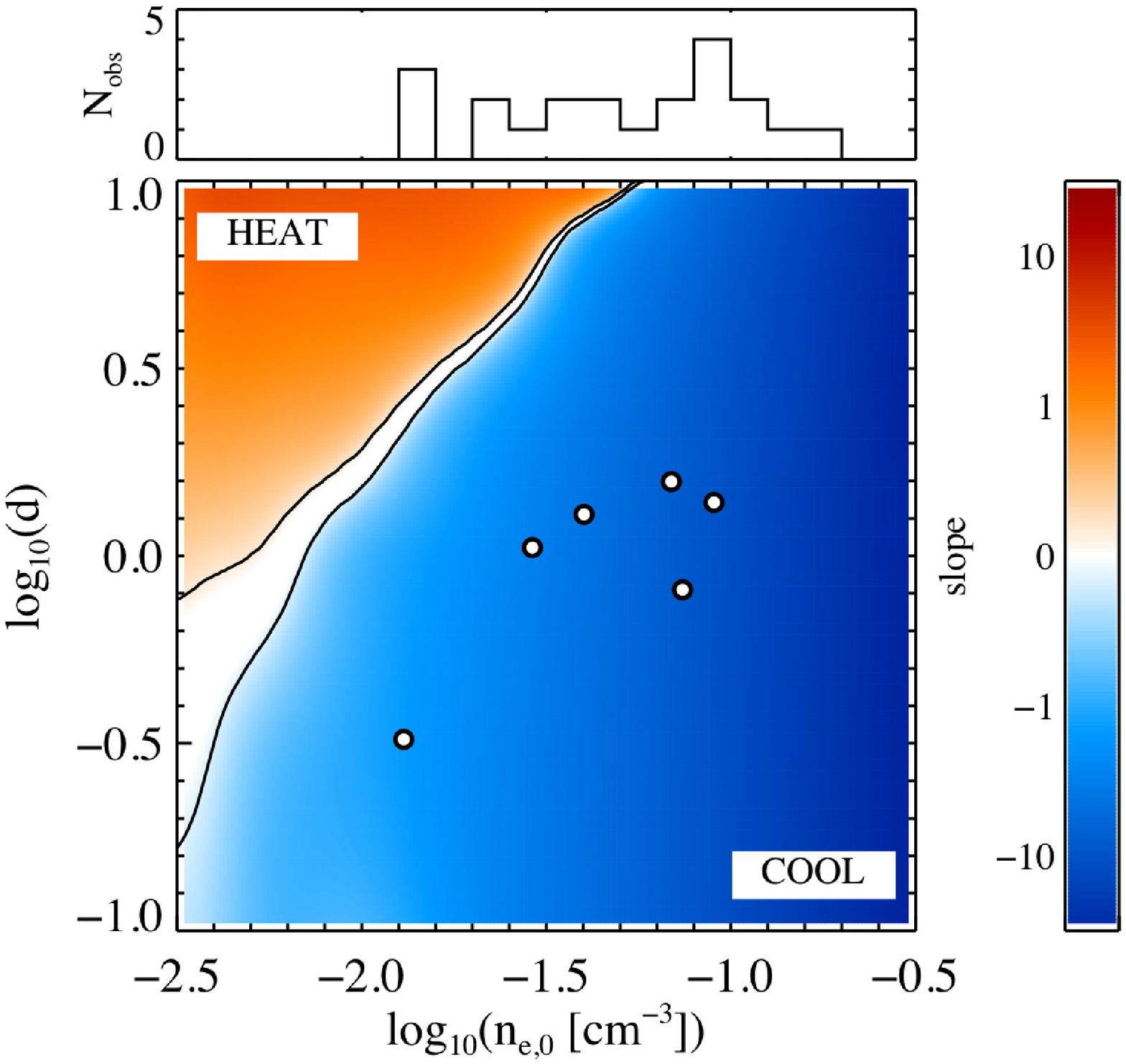}{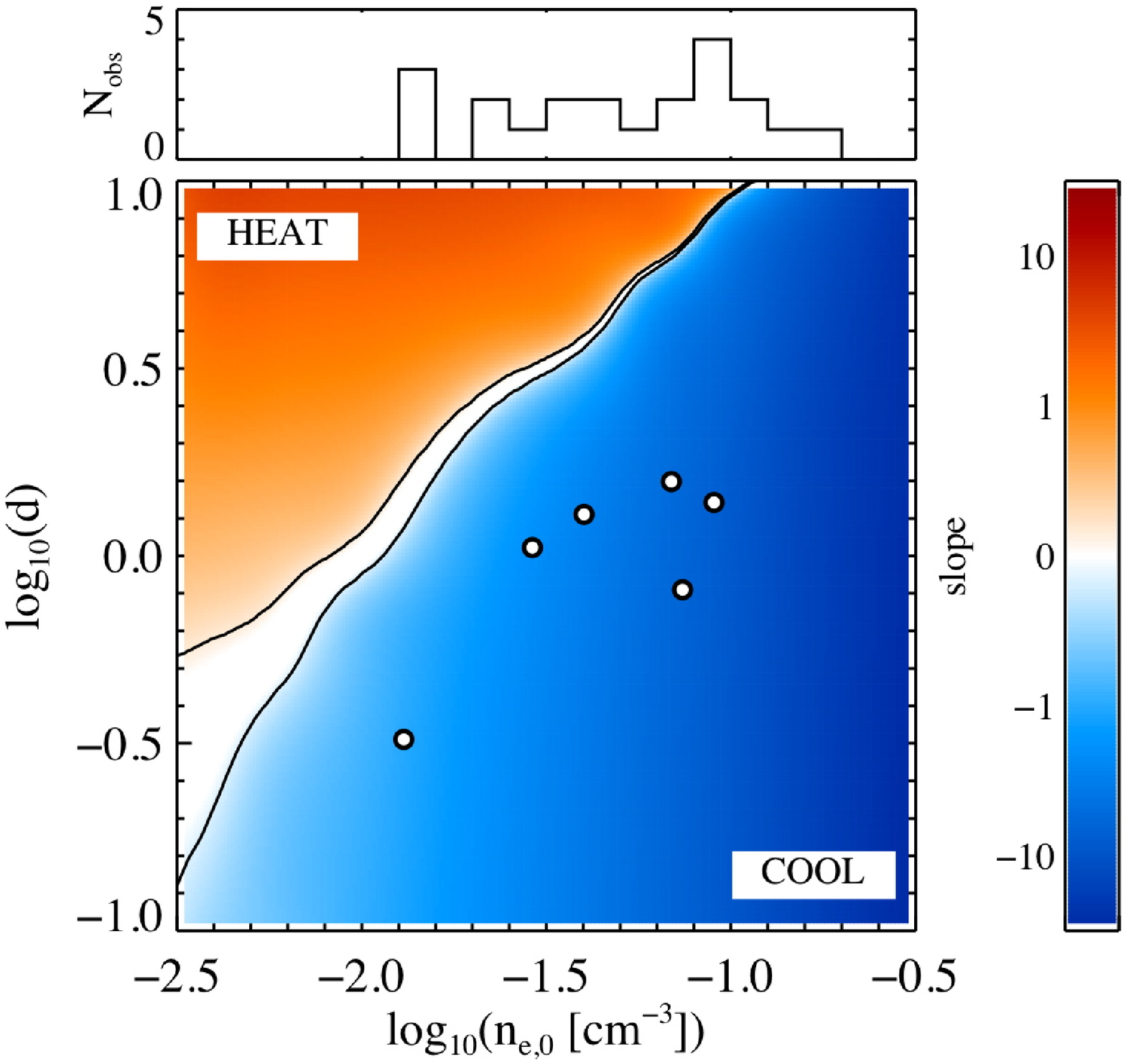}
\vspace{0.5cm}
\caption{Long-term thermal evolution of intracluster media subjected
  to both DF heating and radiative cooling, as a function of initial
  central density, $n_{e,0}$ and normalization of DF heating, $d$
  (cf. Equation \ref{eqn:df}).  Both NCC clusters with $T_{i,NCC}=6$
  keV (\emph{left panel}) and CC clusters with $T_{i,CC}=6$ keV
  (\emph{right panel}) are shown.  The color indicates the slope of
  central temperature vs. time.  Red indicates the region of parameter
  space where the cluster gas catastrophically heats and blue where
  the gas catastrophically cools (i.e. where the slope is positive or
  negative).  The solid black lines indicate the region of parameter
  space where the slope is equal to $\pm0.1$.  Note that while there
  is a domain where heating stably balances cooling, sensitive
  fine-tuning is required for a real cluster to remain in this domain.
  For comparison, the distribution of observed central electron
  densities is included at the top (see $\S$\ref{s:ie}) and the
  estimated location of several Abell clusters are included in the
  $d-n_{e,0}$ plane (\emph{circles}).}
\vspace{0.5cm}
\label{fig:stabdf}
\end{figure*}

As discussed in the Introduction, in the absence of any heating
mechanisms the ICM will cool catastrophically. The time required to
radiate away all of the internal energy of the gas is defined as the
cooling time\footnote{Note that this cooling time differs from the
  conventional cooling time by a factor of
  $\frac{\gamma_g}{\gamma_g-1}$ because the conventional cooling time
  is defined as the time required to cool \emph{isobarically} whereas
  the cooling time in Equation \ref{eqn:tcool} is simply the time
  required to radiate away all of the thermal energy.}:
\noindent
\be
\label{eqn:tcool}
t_{\rm{cool}} \equiv \frac{\rho_g e}{\Lambda} = 5.0\, \bigg(\frac{0.01\, {\rm cm}^{-3}}{n_e}\bigg)  \bigg(\frac{T}{6\, {\rm keV}}\bigg)^{0.5} {\rm Gyr}.
\ee
\noindent
As can be seen from Equation \ref{eqn:tcool}, the central cooling time
for typical clusters is less than a Hubble time.

Figure \ref{fig:rad_run} shows the evolution of the ICM for two
clusters with radiative cooling and no heating.  It is clear that the
clusters cool catastrophically in the absence of any heating.  The top
panel shows the evolution of an initially isothermal cluster with
$T=6$ keV and an initial central electron density $n_{e,0}=0.02$
cm$^{-3}$; the bottom panel shows the evolution of an initially CC
cluster with the same central electron density.  The cooling times for
the NCC and CC clusters shown in the figure are $2.4$ and $1.7$ Gyr,
respectively, based on Equation \ref{eqn:tcool}, and adequately
captures the actual time for catastrophic collapse.

While there are clusters with temperature profiles similar to that
shown in the top panel of Figure \ref{fig:rad_run}, it is important to
note that not only are these temperature profiles transient (in the
sense that the cluster continues to cool on rapid timescales), but the
density profile of this radiation-only cluster (not shown) is nowhere
observed in nature.  Hence the $X$-ray luminosities, which scale as
the gas density squared, would be out of the range of those observed
in nature.  The conclusion that observed clusters must be periodically
heated is in accord with a large body of previous work.  In
particular, $X$-ray spectroscopy has firmly established that the ICM
is not cooling much below $\sim1/3$ of the ambient temperature
\citep{Peterson01, Peterson03, Tamura01}.

The following sections discuss possible heating mechanisms that can
forestall the cooling catastrophe visible in Figure \ref{fig:rad_run}.

\subsection{Radiation \& DF Heating}

\begin{figure}[!t]
\plotone{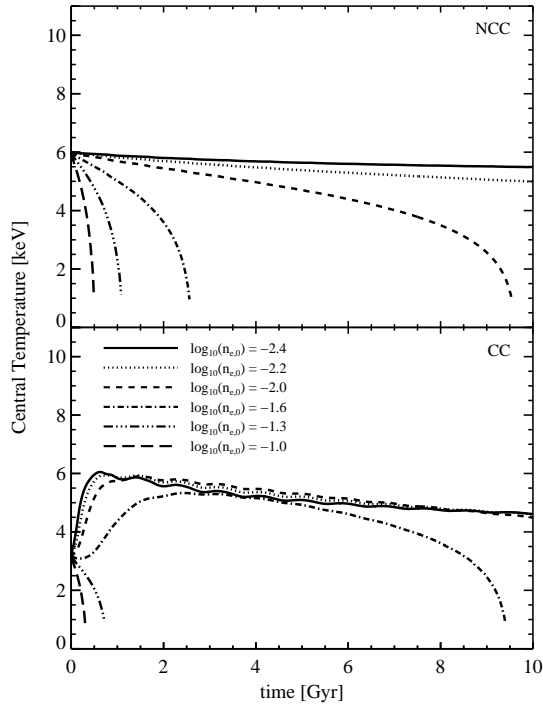}
\vspace{0.5cm}
\caption{Evolution of the central temperature for simulations that
  include both radiation and conduction.  The conduction
  normalization is $f=0.1$ (cf. Equation \ref{eqn:cond}).  \emph{Top
    Panel:} Initially NCC clusters with $T_{i,NCC}=6$ keV.
  \emph{Bottom Panel:} CC clusters with $T_{i,CC}=6$ keV.}
\vspace{0.5cm}
\label{fig:rad_cond}
\end{figure}

\begin{figure}
\plotone{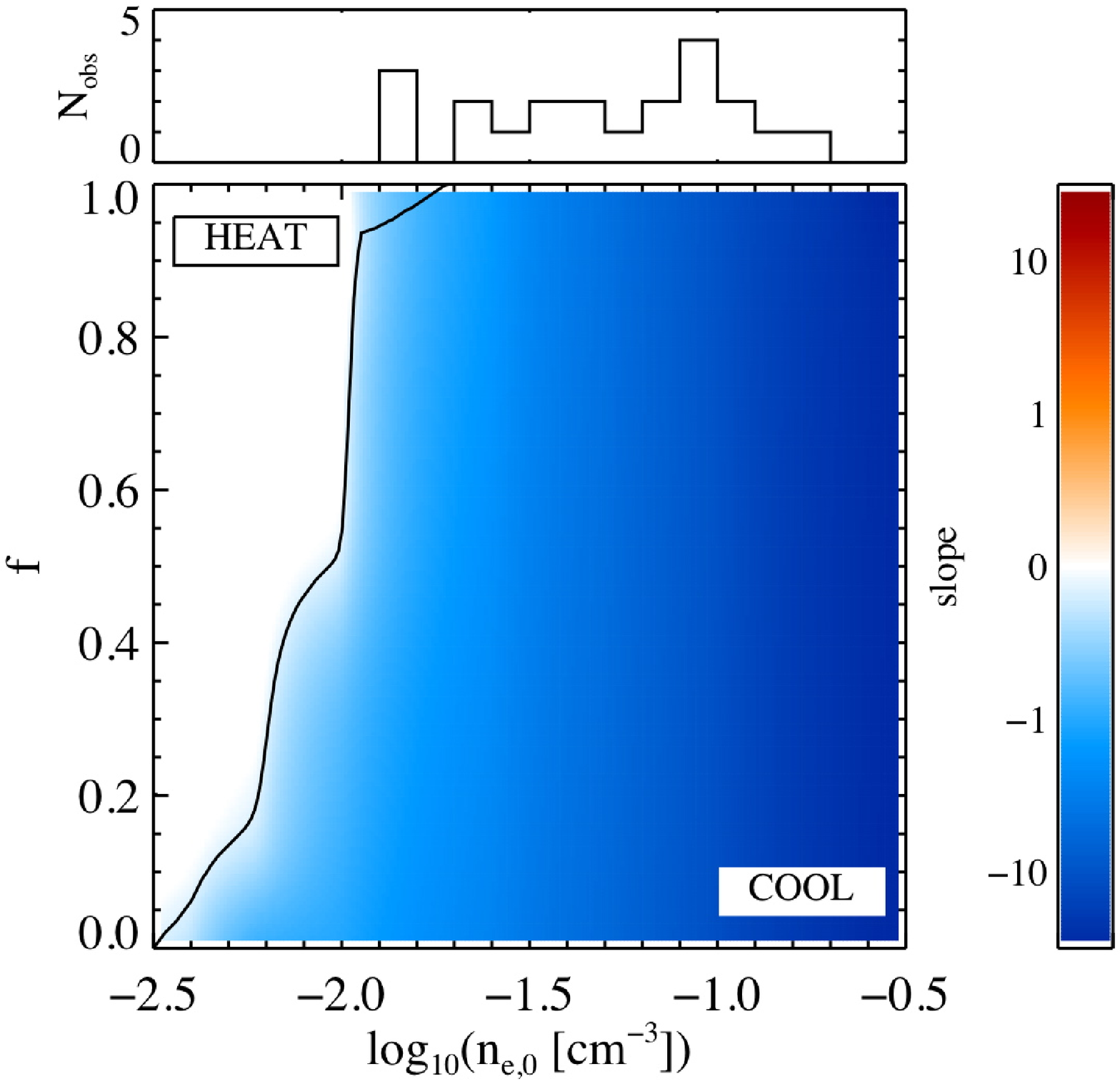}
\plotone{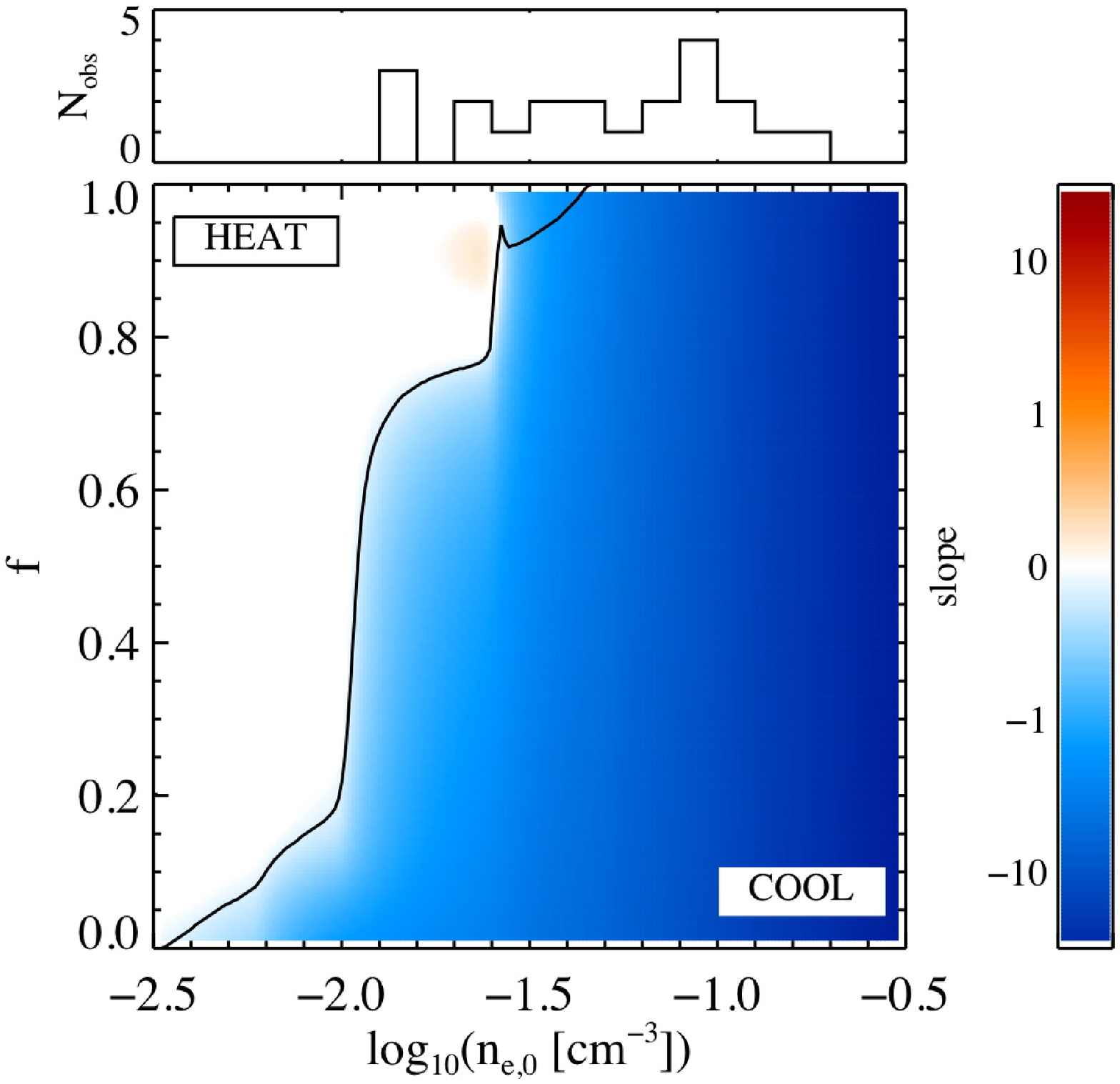}
\vspace{0.5cm}
\caption{Same as Figure \ref{fig:stabdf} except now only thermal
  conduction and radiative cooling are considered, and the clusters
  are initially all NCC.  The normalization of the conductivity, $f$
  (cf. Equation \ref{eqn:cond}), is varied as a function of the
  initial central electron density $n_{e,0}$.  The initial gas
  temperatures are 5 keV (\emph{top panel}) and 8 keV (\emph{bottom
    panel}).}
\vspace{0.5cm}
\label{fig:stabcond}
\end{figure}

As discussed in $\S$\ref{s:df}, there is an enormous source of energy
in the orbital motions of satellite galaxies.  While satellite
galaxies can transfer their orbital energy to the ICM through a
variety of mechanisms, we will focus only on DF heating for
simplicity.  Note however, that other mechanisms capable of
transferring the orbital energy of satellites to the background gas
are expected to scale in a similar way as DF heating (i.e. as $\rho_g$
at fixed temperature).  The discussion that follows thus roughly
encompasses a variety of heating mechanisms related to the motions of
satellite galaxies.

Figure \ref{fig:rad_df} shows the evolution of the ICM central
temperature, $T_c$, as a function of time for initially NCC
(i.e. isothermal; \emph{top panel}) and CC clusters (\emph{bottom
  panel}).  Each panel shows the central temperature evolution for a
range of initial central gas densities.  The normalization of DF
heating shown in the figure is set to $d=1.0$.

Several trends are apparent.  First, it is clear that one can find a
particular set of parameters that leads to approximate thermal balance
(neither a heating nor cooling catastrophe) over a Hubble time.  While
not shown, we note that in the runs where the central temperature does
not change appreciably, neither do the density nor temperature
profiles.  This was first demonstrated by \citet{Kim05}.  In fact, our
fiducial values for DF heating are the same as those in \citet{Kim05}.
The equilibrium NCC model has $T_{i,NCC}=6$ keV and $n_{e,0}=6\times
10^{-3}$ cm$^{-3}$ which is exactly the set of equilibrium parameters
found in \citet{Kim05}, thereby confirming the results of that work.

It is also apparent that DF heating as the sole heating mechanism
leads to serious thermal imbalance for all but a very narrow range of
electron densities, as was also demonstrated by \citet{Kim05}.  This
imbalance arises because at fixed temperature the cooling scales as
$\rho_g^2$ while DF heating scales as $\rho_g$.  Thus at fixed
temperature there can be only one initial density that is in thermal
balance, and that solution will be unstable.

In order to more fully elucidate the thermal properties of the ICM,
Figure \ref{fig:stabdf} shows the long-term thermal evolution of the
ICM as a function of initial central density, $n_{e,0}$ and
normalization of DF heating, $d$, for NCC (\emph{left panel}) and CC
(\emph{right panel}) clusters for a fixed initial temperature profile.
In this figure the slope of the central temperature as a function of
time is color-coded according to the sign and steepness of the slope.
From Figure \ref{fig:rad_df} it is clear that $T_c$ for most clusters
evolves at least quasi-linearly, and hence the slope is a reasonable
metric for the evolution of the ICM.

Figure \ref{fig:stabdf} demonstrates that there is only a narrow range
of thermal balance in the parameter space of $d$ and $n_{e,0}$.  This
highlights the inability of DF heating operating alone to balance
radiative losses in the ICM.  The normalization of DF heating that is
required to balance radiative losses increases in proportion to the
central electron density ($d\propto n_{e,0}$), which makes the DF
heating term (Equation \ref{eqn:df}) effectively scale as $\rho_g^2$.
This is not surprising because the cooling function scales as
$\rho_g^2$, and so thermal balance can be achieved for a range of
densities if the effective DF heating scales in the same way as
cooling.

Nature may in fact provide this required scaling ($d\propto n_{e,0}$).
For example, the number of galaxies per cluster is correlated with the
cluster mass as $N_{\rm gal}\propto M^{0.8}$ \citep{Lin04a}, with
apparently little change at higher redshift \citep{Lin06}, and the
central electron density scales roughly as the cluster mass:
$n_{e,0}\propto M$ \citep[though with large scatter;][]{Vikhlinin06a,
  Zakamska03}.  Since we have incorporated the variation of all free
parameters in Equation \ref{eqn:df} into the single parameter $d$, it
follows that $d\propto N_{\rm gal}$. These additional scalings may
thus in fact produce $d\propto n_{e,0}^\beta$ with $\beta\sim 1$.

In order to investigate this explicitly, we have plotted the
approximate locations of several Abell clusters in the $d-n_{e,0}$
plane in Figure \ref{fig:stabdf}, where the central electron density
is estimated from \citet{Vikhlinin06a} and the number of galaxies,
$N_{\rm gal}$, is provided by the luminosity functions of
\citet{Lin04a}.  We have only included the dependence of $d$ on
$N_{\rm gal}$ when estimating the different DF efficiencies for these
clusters since this scaling is the dominant one.  In other words, for
these clusters we adopt our fiducial values relevant for DF heating
except for $N_{\rm gal}$, which we take from the literature.
Moreover, the intracluster media of these clusters span a range in
temperature \citep[from $\sim2-9$ keV;][]{Lin04b}.  Since the region
of thermal balance in Figure \ref{fig:stabdf} is only a weak function
of temperature, we include all the clusters on the same plot for
simplicity.  A trend of increasing $d$ with increasing $n_{e,0}$ is
evident, but the normalization is a factor of $\sim3$ lower than the
region where DF heating balances cooling.  While a factor of three is
probably within the uncertainty of our estimates of $d$ for these
clusters, what is more important is the narrowness of the region.  Our
estimates of $d$ would not only have to be systematically off by a
factor of $\sim3$, but there would have to be rather extreme
fine-tuning, especially at high $n_{e,0}$, for DF heating to be
generically be able to balance radiative losses.  Moreover, even if
clusters initially fell in this narrow range, DF heating would still
respond to perturbations in $n_{e,0}$ only linearly, while cooling
would respond quadratically, and thus it is unlikely that clusters
would remain in that narrow region for long.

The conclusion drawn from this section is straightforward and echoes
the conclusion in \citet{Kim05}: DF heating, though an important
reservoir of energy, cannot be the sole heating mechanism operating to
offset radiative losses in the ICM.  This is not because the energy
available is insufficient but rather because there is only a narrow
range of parameter space where DF heating can stably balance radiative
cooling.

\subsection{Radiation \& Conduction}\label{s:cond}

Thermal conduction can in principle transport heat from the abundant
thermal reservoir of the outer cluster gas to the inner cooling parts.
Invoking conduction as a means to balance radiative losses in the ICM
appears however to require a fair degree of fine tuning of the
conductivity \citep{Zakamska03, Kim03a, Pope06, Guo07}.  As
demonstrated below, this problem is particularly acute in CC clusters
where the {\it observed} temperature profile drops precipitously in
the inner parts --- too little conduction leads to a cooling
catastrophe, while too much conduction tends to produce an isothermal
core.

These issues are demonstrated graphically in Figure \ref{fig:rad_cond}
where the evolution of the central temperature is plotted for both NCC
(\emph{top panel}) and CC (\emph{bottom panel}) clusters, for a
conductivity normalization $f=0.1$.  Notice that the evolution is
qualitatively different for NCC and CC clusters.

For NCC clusters, which are initially isothermal, only clusters with
very low gas densities are marginally stable, while progressively more
dense clusters quickly runaway.  These trends can be understood with
the aid of the conduction time:
\noindent
\be
t_{\rm cond} \equiv \frac{0.4}{f}\, \bigg(\frac{n_e}{0.01\, {\rm cm}^{-3}}\bigg) \bigg(\frac{\lambda}{100\,{\rm kpc}}\bigg)^2 \bigg(\frac{6\, {\rm keV}}{T}\bigg)^{2.5} {\rm Gyr},
\label{eqn:tcond}
\ee
\noindent
where $f$ is the conductivity normalization and $\lambda$ is the
length scale over which temperature gradients are washed out within a
conduction time \citep[the inner region where the cooling time is
shorter than a Hubble time is generally of order 100
kpc;][]{Sanderson06}.  It is clear that larger densities lead to
longer conduction times\footnote{The conduction time depends on
  density because, while conduction transports energy at fixed
  density, energy scales as $n_e T$, so that the conduction time
  scales as $n_e$.}, and that for $f=0.1$, densities much greater than
$n_{e,0}\sim 10^{-2} $ cm$^{-3}$ result in conduction times
approaching the Hubble time, and much longer than the cooling time.
For these high densities, conduction cannot forestall runaway cooling,
as seen in Figure \ref{fig:rad_cond}.  Thermal conduction alone cannot
therefore stably offset radiative losses in the intracluster media of
NCC clusters.

The situation is somewhat more complex in CC clusters because there
the initial temperature profile drops by a factor of $\sim2-3$ within
the inner $\sim 100$ kpc.  The bottom panel of Figure
\ref{fig:rad_cond} shows the central temperature evolution for these
clusters.  For clusters with short conduction times, the inner
(cooler) region isothermalizes to the temperature of the outer cluster
region ($\sim6$ keV for these runs) before radiative effects become
important.  Once isothermal, the clusters then evolve more like the
NCC clusters (\emph{top panel}, Figure \ref{fig:rad_cond}).  CC
clusters that are too dense have conduction times that are longer than
the cooling times, and they thus cool catastrophically.  These trends
persist for the full range of conductivities explored herein
($0<f<1$).

The trends evident for NCC clusters in the top panel of Figure
\ref{fig:rad_cond} are shown for a wide range in conductivity
normalizations and central electron densities in Figure
\ref{fig:stabcond}.  We have simulated NCC clusters with an initial
temperature of 5 keV ({\it top panel}) and 8 keV ({\it bottom panel})
in order to demonstrate explicitly that our results are relatively
insensitive to the ICM temperature.  Unlike Figure \ref{fig:stabdf},
NCC clusters can only cool or remain in thermal balance, since
conduction acting on an isothermal gas cannot produce a steadily
increasing central temperature with time.  It is clear that increasing
the conductivity results in stable intracluster media over a wider
range of initial central electron densities, but the range where
cooling balances heating increases only mildly as the conduction
normalization increases from $f\sim0.2$ to $f\sim1.0$.

The DF heating and conduction fine tuning problems are thus somewhat
different. In the former case, too much DF heating will result in
runaway heating, while in the latter, too much conduction simply
isothermalizes the core.  This is not acceptable because the majority
of clusters show a factor of $\sim2-3$ temperature drop within their
cores \citep[e.g.][]{Sanderson06, Vikhlinin06a}.  The implications are
however the same: neither mechanism is generically able to reproduce
observed temperature profiles.

It thus appears that previously proposed models for the ICM that
construct clusters initially in thermal balance with conduction as the
only heating source \citep{Zakamska03, Kim03a, Guo07} constitute a
very narrow regime of parameter space and should thus not be
considered realistic solutions to the cooling flow problem, unless a
physical explanation for the fine-tuning is provided \citep[see also
the discussion in][]{Guo07}.

\subsection{Radiation, DF Heating, \& Conduction}

In this section the thermal balance of the ICM is explored when both
DF heating and conduction are operating as heating mechanisms.

\begin{figure}[!t]
\plotone{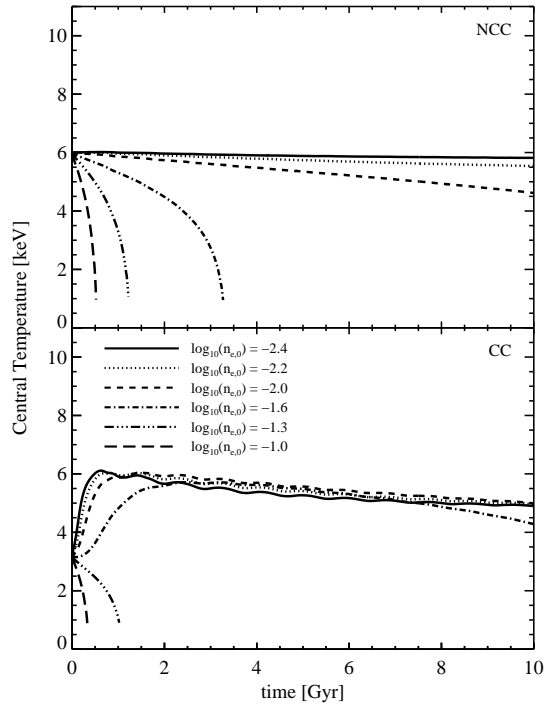}
\vspace{0.5cm}
\caption{Evolution of the central temperature for simulations that
  include radiation, DF heating, and conduction.  The DF heating and
  conductivity normalizations are $d=1.0$ and $f=0.1$, respectively.
  \emph{Top Panel:} Initially NCC clusters.  \emph{Bottom Panel:}
  Initially CC clusters.}
\vspace{0.5cm}
\label{fig:rcd}
\end{figure}

Figure \ref{fig:rcd} plots the evolution of the central temperature
for NCC and CC clusters with normalization of DF and conduction set to
$d=1.0$ and $f=0.1$.  Comparing this figure to Figures
\ref{fig:rad_cond} and \ref{fig:rad_df} highlights the stabilizing
effects of the combination of DF heating and conduction.  Note however
that the evolved CC clusters are plagued by the same issues discussed
in $\S$\ref{s:cond}, namely that while the combination of DF and
conduction can maintain CC clusters in thermal balance, the evolved
clusters do not preserve the factor of $\sim2-3$ drop in the
temperature profile characteristic of observed CC clusters.  In other
words, these CC clusters that do not catastrophically cool instead
turn into NCC clusters, and thus these runs cannot explain the
existence of observed CC clusters.

\begin{figure}[t]
\plotone{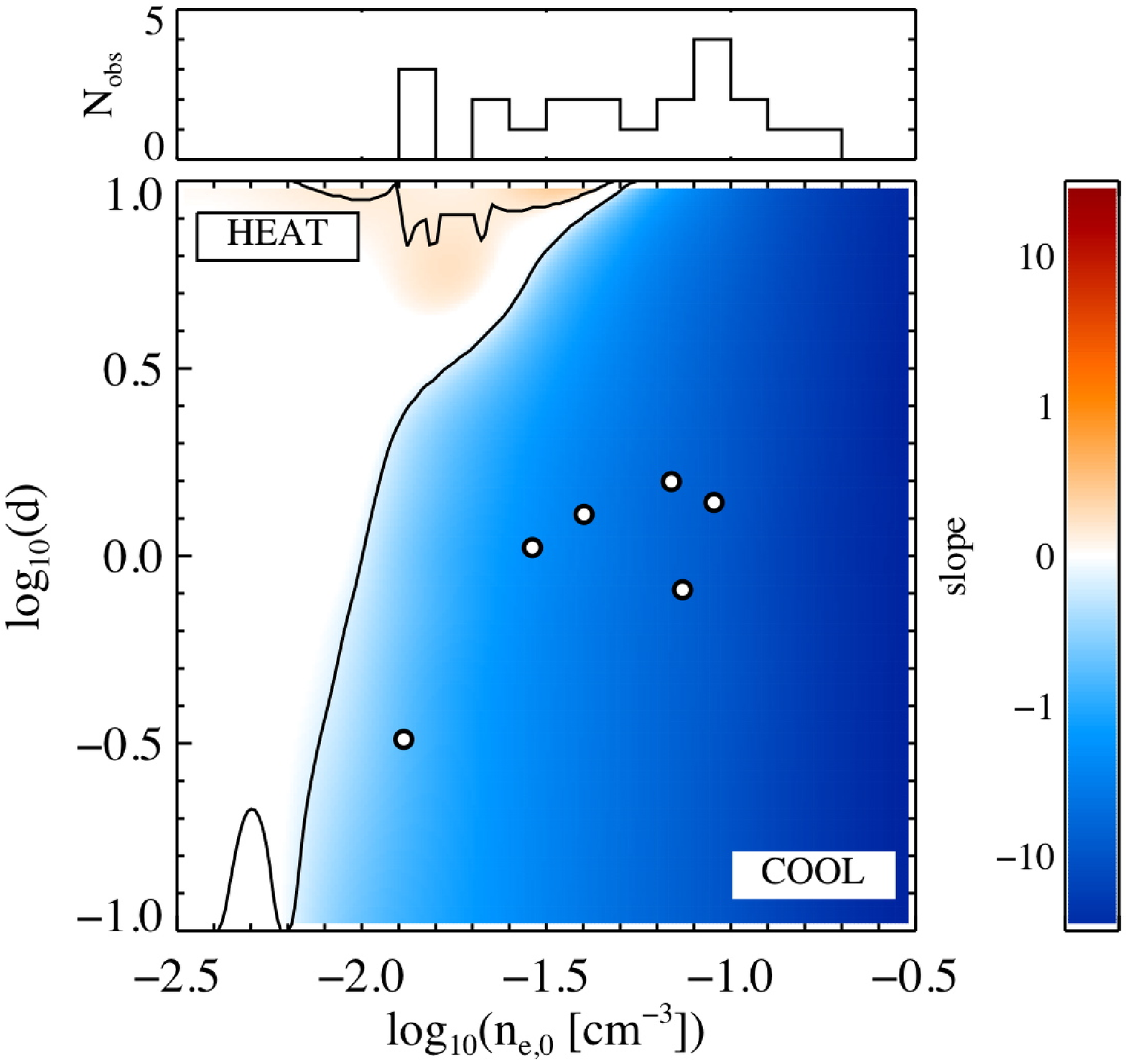}
\plotone{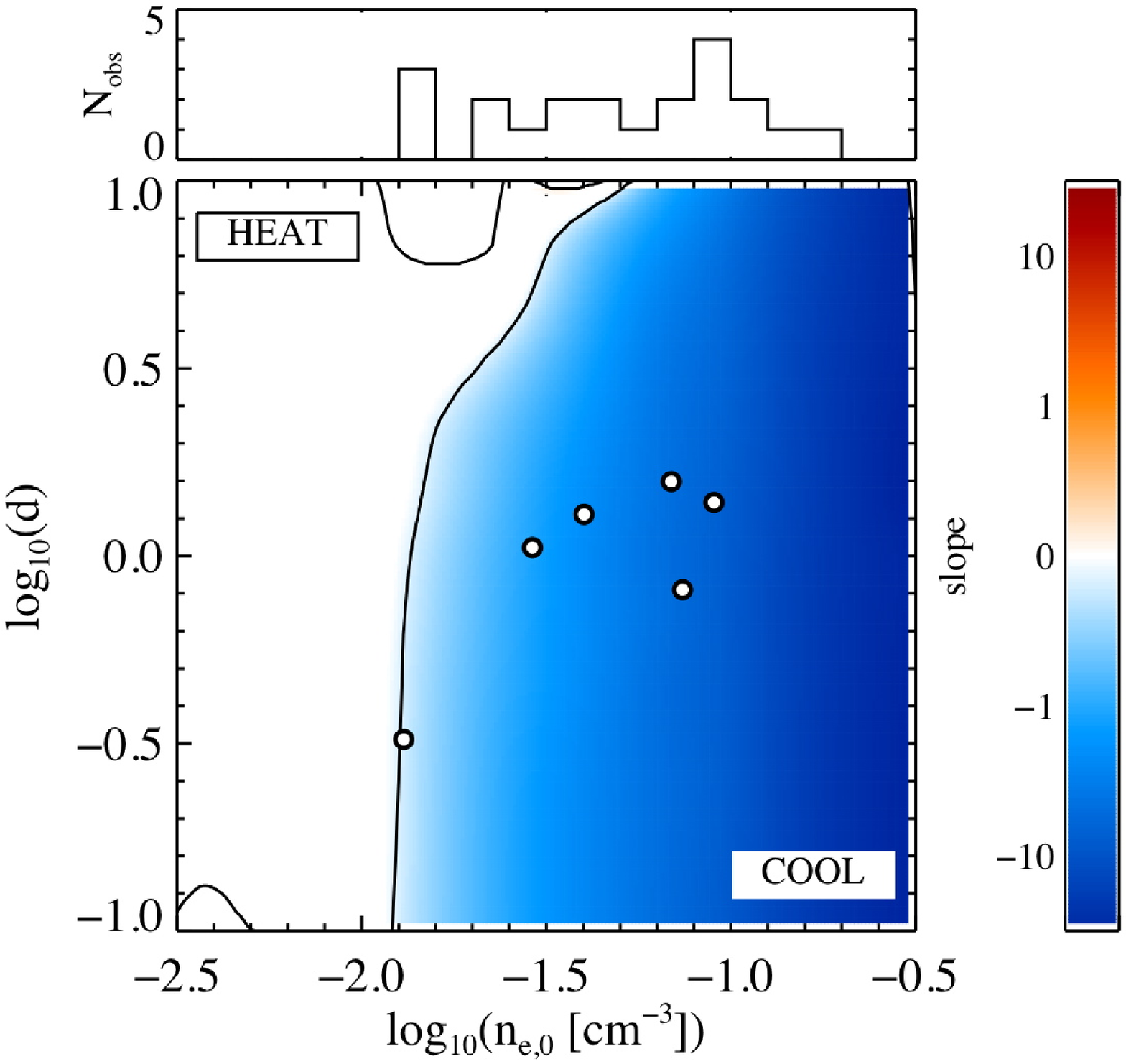}
\vspace{0.5cm}
\caption{Same as Figure \ref{fig:stabdf}, now with DF heating, thermal
  conduction, and radiative cooling, for NCC clusters with
  $T_{i,NCC}=6$ keV.  \emph{Top Panel:} NCC clusters with the
  conduction normalization set to $f=0.1$.  \emph{Bottom Panel:} NCC
  clusters with $f=0.5$.  Note that even a small amount of conduction
  makes the clusters much more thermally stable than DF heating alone
  over a much wider range of central gas densities.  As in Figure
  \ref{fig:stabdf} the approximate locations of several Abell clusters
  are included (\emph{circles}).}
\vspace{0.5cm}
\label{fig:rad_cond_df}
\end{figure}

The increased thermal balance when combining DF heating with
conduction is manifest in Figure \ref{fig:rad_cond_df}, which shows
the evolution of the central temperature of the ICM for initially NCC
clusters with $T_{i,NCC}=6$ keV in the parameter space of initial
central electron density and normalization of DF heating.  This figure
shows the effect of including conduction with normalization $f=0.1$
(\emph{top panel}) and $f=0.5$ (\emph{bottom panel}).

Upon comparing this figure to Figures \ref{fig:stabdf} and
\ref{fig:stabcond}, it is clear that the combination of DF heating and
conduction makes the ICM of NCC clusters considerably more thermally
balanced than either mechanism acting alone.  Indeed, for initial
central electron densities $n_{e,0}\lesssim 0.02 $ cm$^{-3}$ the
ICM is thermally balanced for any plausible value of the DF heating
normalization, and for reasonable, not fine-tuned, choices of
conductivity.

However, it is still the case that for $n_{e,0}\gtrsim 0.02 $
cm$^{-3}$, no plausible amount of DF heating can offset the cooling
catastrophe.  Increasing the amount of conduction to $f=1.0$ expands
the zone of thermal balance to the right in Figure
\ref{fig:rad_cond_df} only marginally.  While some observed NCC
clusters have a central density lying within the region of thermal
balance in Figure \ref{fig:rad_cond_df}, it is clear that for many
observed clusters processes in addition to DF heating and conduction
are at work in order to effectively balance radiative cooling.

These conclusions are rather insensitive to the initial temperature of
the NCC and CC clusters.  This might be expected because the cooling
function scales as $T^{1/2}$ while DF heating scales at $T^{-1/2+p}$
with $p\sim0$ \citep{Kim05}.  For example, decreasing the initial
temperature in Figure \ref{fig:rad_cond_df} by a factor of two results
in somewhat more thermally balanced clusters in the high density
regime.  However, the increased zone of thermal balance is narrow at
high density and hence rather fine tuning would be required for an
observed cluster to remain there.  Moreover, observed clusters with
larger central densities have higher rather than lower temperatures,
and so the problem of generating equilibrium clusters is only
exacerbated if the observed temperature-density relation is
considered.

In sum, while it appears that the combination of DF and conduction can
produce intracluster media in thermal balance for a modest range of
central electron densities for NCC clusters, neither NCC clusters with
moderately high central electron densities nor CC clusters of any
density can be kept in thermal balance for cosmological timescales
with these mechanisms alone.

\subsection{The Impact of a Relativistic Fluid}

A relativistic fluid may be present in observed clusters, perhaps in
the form of cosmic rays \citep{Pfrommer04, Sanders05, Dunn06,
  Sanders07, Werner07, Nakar07}.  We show in the appendix that the
addition of a small amount of relativistic pressure increases the
thermal stability of a gas already in thermal balance (i.e. the growth
of thermal perturbations is damped for intracluster media where the
net heating of a parcel of gas is zero).  In this section we explore
the impact of a relativistic fluid on the thermal {\it balance} of the
ICM.

As discussed in $\S$\ref{s:prel}, a relativistic fluid is included in
our simulations by requiring that its pressure be a fixed fraction of
the thermal pressure initially.  The relativistic fluid is assumed to
be massless and perfectly dynamically coupled to the gas.  Its
subsequent adiabatic evolution is then governed by energy
conservation.  Recall that in our treatment the relativistic fluid
provides no additional thermal energy to the system.  The purpose of
this experiment is thus to ascertain the potential importance of a
relativistic fluid on the hydrodynamics of the ICM, not the
thermodynamics.

We have run a series of simulations that span the range of parameter
space shown in Figure \ref{fig:stabdf}, i.e. for a range in initial
central electron densities and DF heating normalizations.  In addition
to varying these two variables, we have also varied the initial
fraction of relativistic pressure, $\alpha$, from 10\% to 40\% of the
total pressure, which is a fraction consistent with observations
\citep{Pfrommer04, Sanders05, Dunn06, Sanders07, Werner07, Nakar07}.
We find that the addition of relativistic pressure has a completely
negligible effect on the thermal balance of the gas for the entire
range of parameter space we have explored.

At first glance this is somewhat surprising since a thermal stability
analysis indicates that a small amount of relativistic pressure should
increase markedly the stability of a gas already in thermal balance
\citep[][see also $\S$\ref{s:append}]{Cen05}.  The key difference is
that stability analyses assume the gas to be in thermal balance, while
the simulations we have run are almost never set initially in thermal
balance.  In the former case the classic \citet{Field65} instability
(described in the Introduction) is suppressed because slightly
over-dense fluid elements, which thus cool rapidly and hence lose
thermal pressure support, contract less than they would otherwise
thanks to the additional pressure support provided by the relativistic
fluid.  However, in the latter case, where the gas in not in thermal
balance, the additional pressure support provided by the relativistic
fluid is immaterial because the thermal runaway is caused by a more
serious imbalance between heating and cooling, which the relativistic
fluid, providing only pressure support and not energy, cannot remedy.

\section{Discussion}
\label{sec:disc}

The results of the preceding section imply that both thermal
conduction and DF heating are potentially significant heating
processes in the ICM.  In fact, for large regions of parameter space,
these processes provide more than enough energy to offset radiative
losses.  However, we have found that they can neither produce nor
maintain intracluster media in thermal balance over cosmological
time-scales.  In particular, these processes tend to generate either
runaway cooling or runaway heating, in both cases producing
temperature and density profiles of the ICM nowhere near those
observed in nature \citep{Sanderson06, Vikhlinin06a}.  These
conclusions expand upon those of \citet{Brighenti02} who found that
none of a large range of possible steady-state heating mechanisms
could generically reproduce the observed properties of the ICM.  For
the case of DF heating, these results are driven largely by the fact
that radiation losses scale as the gas density squared, while DF
heating scales linearly with the density.  Since observed intracluster
media exhibit a wide range of densities, fine-tuning is required to
prevent thermal runaways when DF is the dominant heating mechanism.

These conclusions have several implications, two of which we now
discuss in detail.  First, in light of our results from one
dimensional simulations, we discuss the continued necessity of
high-resolution, fully three-dimensional, cosmologically embedded
hydrodynamic simulations for understanding the thermodynamics of the
ICM.  Second, we discuss a heating mechanism that may plausibly
generate and maintain the properties of observed intracluster media,
since none of the mechanisms explored herein appear capable of doing
so.

\subsection{The Cosmological Context}

Our approach has been to simulate in one dimension the evolution of
the ICM when subjected to various heating mechanisms and radiative
cooling in static, isolated clusters, where only the ICM was evolved.
In the real Universe, however, clusters do not appear to evolve in
isolation.  Indeed, cosmological cold dark matter $N$-body simulations
demonstrate that cluster-sized dark matter halos ($M\gtrsim10^{14}\,
M_\Sun$) continually accrete additional halos with a range of masses,
many of which likely host galaxies.  The mass growth rate is
time-dependent, slowing at late times \citep[e.g.][]{Wechsler02}.  The
cosmological context within which clusters evolve may in fact play a
significant role in their thermal history, as suggested by recent
hydrodynamic simulations \citep[e.g.][]{Motl04, Burns07, Nagai07,
  McCarthy07a}.

There are a variety of ways in which the cosmological setting can
provide additional heating mechanisms.  A generic class of such
mechanisms may be called ``gravitational'' in the sense that the
energy available for heating ultimately comes from the gravitational
potential energy of infalling material.  DF heating is one example of
this class, but there are others.  In particular, if small gaseous
clumps are accreted, they may transfer their gravitational energy to
the ICM via ram pressure drag and local shocks
\citep[e.g.][]{Murray04, McCarthy07a, Khochfar07, Dekel07}.  Major
mergers can significantly mix the ICM, effectively transporting the
abundant reservoir of energy in the cluster outskirts to the cooler
inner regions \citep[e.g.][]{Burns07}.

Recent observations have suggested that the fraction of CC clusters is
increasing with time since at least $z\sim1$ \citep{Vikhlinin06b,
  Ohara07}.  If gravitational heating plays a significant role in the
thermal history of the ICM, then one may expect that as the accretion
rate slows at late times \citep{Wechsler02}, radiative losses would
become increasingly dominant, in at least qualitative agreement with
these recent observations.  More generally, it is clear that the
time-evolution of the abundance of observed CC clusters will provide
strong, unique constraints on the importance of various heating
mechanisms of the ICM.

While it may be appealing to invoke major mergers or other strong
mixing processes as a means to transfer energy into the inner cooling
regions of clusters, such processes have potential draw-backs.  The
most serious is that observed CC clusters have strong metallicity
gradients within the cooling region \citep{DeGrandi01}.  If the metals
are produced by type Ia SNe within the central galaxy then the
timescale for generating the observed gradient is $\sim5$ Gyr
\citep{Bohringer04}, which is much longer than the cooling time of the
ICM in most observed clusters \citep{Sanderson06}.  Sedimentation may
also contribute to the metallicity gradient on relevant timescales,
but its relevance is much more uncertain because it can be highly
suppressed in the presence of even modest magnetic fields
\citep[e.g.][]{Fabian77b}.  If mergers are capable of mixing the ICM
such that the inner cooling region is effectively heated, it seems
plausible that they may also destroy the metallicity gradient.
Detailed numerical simulations are required to address these issues.

It is clear that 3D cosmologically embedded simulations are preferable
to 1D simulations of the type presented herein because there are
potentially relevant physical processes that cannot be adequately
captured in the latter approach.  While we have highlighted the
potential importance of the cosmological context, it is also true that
certain physical processes cannot be captured simply because our
simulations are in 1D.  For example, processes such as convection and
the Rayleigh-Taylor instability cannot be captured in a 1D simulation.
Additional insight into the cooling flow problem can thus be gained by
simply extending the types of simulations we have run into 3D.

In spite of these fact, the motivations for focusing on 1D simulations
are several-fold.  First, cosmologically embedded simulations are very
time-consuming --- a parameter space study similar to what we have
presented in this work is not currently possible with such
simulations.  Second, DF is a resonant process \citep{Tremaine84}, and
so the particle requirement to adequately resolve this phenomenon may
be considerably higher than what is achievable in the current
generation of cosmologically embedded hydrodynamic simulations.  It is
not sufficient to resolve well the dark matter halo and subhalos.
Since much of the DF heating is caused by the inspiralling of the much
smaller stellar systems, a resolution small compared to the $\sim10$
kpc half-light radii of giant ellipticals is required.  Thus, while
our approach to DF heating has been idealized, it was necessary to
explore its possible effects in this parameterized manner because it
is not clear that current 3D simulations are adequately resolving this
phenomenon \citep[see discussion in][]{Faltenbacher05, Naab07}.
Finally, even with sufficient resolution, there are a number of
physical effects that we have demonstrated are potentially relevant
for the thermodynamics of the ICM that are not included in the
majority of cosmologically embedded hydrodynamic simulations.  Such
phenomena include thermal conduction, which ultimately requires the
inclusion of magnetic fields, type Ia supernovae energy injection,
which is included in some but not all current simulations, and energy
injection from accretion onto a central black-hole, which we now
discuss in detail.

\subsection{A Self-Regulating Heating Mechanism}

In the present work we have focused on heating processes that are not
generally thought of as ``feedback'' mechanisms.  While DF heating is
in principal a feedback mechanism since the heating becomes
inefficient at very and low high Mach numbers, in practice this
feedback is too weak to prevent runaway heating or cooling.  The
preceding discussion of the importance of the cosmological context
suggests that additional gravitational energy is available that can
contribute to the heating of the ICM.  However, the incredibly short
cooling times of many CC clusters ($\ll 1$ Gyr) indicates that heating
from without, via gravitational processes, must still be fine-tuned in
order to prevent runaway cooling.  Heating from within, via the
release of energy from accretion onto a black hole, is in many ways
more appealing because such a process is explicitly self-regulating.

Black hole accretion-mediated feedback mechanisms
\citep[e.g][]{Ciotti01, Ruszkowski02, Kaiser03, Guo07, Ciotti07} are
qualitatively different from the mechanisms we have explored.  These
self-regulating feedback processes are thought to work schematically
as follows.  As the gas begins to cool in the inner regions, its
thermal pressure no longer provides sufficient support to the
overlying material, and thus matter flows inward and eventually
accretes onto the central black hole.  The mechanical energy provided
by AGN activity, which is proportional to the rate of mass infall,
then heats up the surrounding gas until the thermal pressure, which is
now increasing due to the energy injection, halts the inward flow of
matter, thereby diminishing the accretion-driven energy injection.
The cycle then begins again in a regulatory fashion, and the ICM thus
neither heats nor cools catastrophically.  It is this regulatory
feedback processes which seems most promising in explaining the
properties of observed intracluster media because the relevant
physical parameters need not be tuned to any particular values for
heating to balance cooling generically \citep[see e.g.][]{Guo07}.

A serious constraint on AGN-related feedback is that it may destroy
the observed metallicity gradient in CC clusters if the energy
deposition at the cluster center is sufficient to drive convection.
Recently, \citet{Voit05} have shown that AGN feedback can be effective
at balancing radiative cooling \emph{and} maintaining the observed
metallicity gradient if the AGN outbursts are rather gentle, occur
every $\sim10^8$ years and last in duration for $\sim10^7$ years.  It
remains to be seen if this proposal can be confirmed with both
numerical simulations and, ultimately, observations.

While AGN heating may be a promising candidate at generating and
maintaining intracluster media in thermal balance over cosmological
time-scales, simulations of the ICM which include it do so only in a
rather simplistic fashion, in part because a detailed understanding of
the physical processes involved in black hole accretion currently
eludes us.  The way in which energy is transfered from the black hole
to the surrounding ICM is also obscure.  Until these and related
issues are better understood, we should continue to seek out other
potential self-regulating heating processes that could potentially be
relevant for the thermodynamics of the ICM.  Nonetheless, varied
observations of sound waves, bubbles, and strong radio emission
\citep{Birzan04, Best05, Fabian06}, predominantly in CC clusters, in
addition to their very short cooling times \citep{Sanderson06},
strongly suggests that AGN-related activity plays an important role in
the thermodynamics of the ICM.

\section{Summary}
\label{sec:sum}

We have presented the results from a series of 1D simulations aimed at
understanding the importance of type Ia supernovae heating, thermal
conduction and DF heating (due to the orbital motions of satellite
galaxies) on the thermal properties of the intracluster media of
clusters.  Both initially non-cooling core (NCC; i.e. isothermal) and
cooling core (CC) clusters were simulated, for a wide range of initial
central electron densities and gas temperatures.

Marginalizing over the uncertain efficiencies of DF heating and
thermal conduction, it is clear that only NCC clusters with central
electron densities $n_{e,0}\lesssim 0.02$ cm$^{-3}$ can be maintained
in thermal balance over a Hubble time if both DF and conduction
operate; neither mechanism alone can generate generically stable ICM
at these densities.  At higher densities no reasonable amount of
conduction or DF heating can prevent runaway cooling in NCC clusters.

The failure of the combination of conduction and DF heating at either
generating or maintaining observed intracluster media is more
pronounced for CC clusters, which have observed temperature profiles
that decline by a factor of $\sim2-3$ from the outer to inner regions.
This temperature drop is extremely difficult to maintain in the face
of thermal conduction in our simulations, because thermal conduction
acts to erase temperature gradients.  In fact, of the
$\mathcal{O}(10^3)$ simulations run with a wide range of DF heating
and conduction efficiencies and initial gas densities, none generated
stable CC cluster profiles.  Since CC clusters constitute $\sim70$\%
of observed clusters \citep[e.g.][]{Peres98}, we regard this failure
as strong evidence that other heating processes in the ICM must be at
work besides conduction and DF heating.

Our results demonstrate that there are numerous energy reservoirs
capable of supplying enough energy to offset radiative losses.  The
crux of the cooling flow problem therefore lies not in finding one or
more mechanisms capable of providing enough energy to the ICM, but
rather in finding one or more mechanisms that can supply the energy in
a way that maintains thermal equilibrium in the ICM over cosmological
time-scales.  Only low-density NCC clusters can be maintained with DF
heating and conduction.  The processes explored herein are not
manifestly self-regulating, and they thus require fine-tuning in order
to generate the observed properties of CC and high-density NCC
intracluster media generically. For these types of observed
intracluster media, it seems likely that an explicitly self-regulatory
feedback process such as black hole accretion-powered energy injection
(i.e. AGN feedback) is required.

\acknowledgments 

CC gratefully acknowledges Jim Stone, Anatoly Spitkovsky, and Ian
Parrish for extensive help and support with extensive numerical
issues, and Andrey Kravtsov for many fruitful discussions.  We thank
Paul Bode, Woong-Tae Kim, Andrey Kravtsov, and Jim Stone for helpful
comments on an earlier draft.



\begin{appendix}

\section{Thermal Stability for Generic Heating Mechanisms In the Presence of a Relativistic Fluid}
\label{s:append}

The generalized Field criterion \citep{Field65, Balbus86} states that
a gas is thermally unstable to isobaric perturbations if:
\noindent
\be
\frac{\partial( \mathcal{L}/T)}{\partial T}\bigg|_P < 0,
\ee
\noindent
where $\mathcal{L}$ is the net loss function per unit mass defined
such that $\rho_g \mathcal{L} = \Lambda-\Gamma$ where $\Lambda$ and
$\Gamma$ are the cooling and heating rates per unit volume.  In what
follows the subscript ``$g$'', denoting gaseous quantities, will be
omitted for brevity.

The following identity holds for a general gas in thermal equilibrium ($\rho \mathcal{L}=0$):
\be
\frac{\partial( \mathcal{L}/T)}{\partial T}\bigg|_P = \frac{1}{\rho T}\bigg[ \frac{\partial(\rho \mathcal{L})}{\partial T}\bigg|_\rho + \frac{\partial \rho}{\partial T}\bigg|_P \frac{\partial(\rho \mathcal{L})}{\partial \rho}\bigg|_T\bigg].
\label{eqn:field}
\ee
The first and third derivatives on the right-hand side are determined by the relevant heating and cooling mechanisms while the second is set by the various sources of pressure support.  The following identity will be useful:
\be
\frac{\partial \rho}{\partial T}\bigg|_P = -\frac{\partial P}{\partial T}\bigg|_\rho \,\bigg/\, \frac{\partial P}{\partial \rho}\bigg|_T.
\ee
\noindent
Assume that the pressure support is provided by both the thermal
pressure of the gas and a relativistic fluid:
\be
P = P_{\rm th} + P_{\rm rel} = K_1 \rho T + P_{\rm rel}(\rho),
\ee
where $K_1$ is a constant and $P_{\rm rel}$ is a function only of density.  It then follows that
\be
\frac{\partial P}{\partial T}\bigg|_\rho = K_1\rho,
\ee
and
\be
\frac{\partial P}{\partial \rho}\bigg|_T = K_1 T + \frac{\partial P_{\rm rel}}{\partial \rho}\bigg|_T,
\ee
and hence:
\be
\frac{\partial \rho}{\partial T}\bigg|_P =\frac{ K_1\rho}{K_1 T + \frac{\partial P_{\rm rel}}{\partial \rho}\big|_T} = \frac{\rho}{T}\frac{P_{\rm th}}{P_{\rm th}+\rho \frac{\partial P_{\rm rel}}{\partial \rho}\big|_T } =  \frac{\rho}{T}\frac{\alpha'}{\alpha' + \frac{\partial{\rm ln}P_{\rm rel}}{\partial {\rm ln} \rho}\big|_T},
\label{eqn:ptot}
\ee
\noindent
where in the last equality we have defined $\alpha'\equiv P_{\rm th} / P_{\rm
  rel}$.

Now assume that cooling is dominated by thermal Bremsstrahlung radiation and that the heating has a simply power-law dependence on $\rho$ and $T$.  Then:
\be
\rho \mathcal{L} = \Lambda_0 \rho^2 T^{1/2} - \Gamma_0 \rho^\beta T^{-1/2+\delta},
\ee
\noindent
where $\Lambda_0$ and $\Gamma_0$ are constants and $\beta$ and
$\delta$ are the power-law indices for heating, defined in a way that
will prove useful below.  After some algebra, and using the fact that
$\rho \mathcal{L}=0$, we have
\noindent
\be
\frac{\partial(\rho \mathcal{L})}{\partial T}\bigg|_\rho = (1-\delta)\Lambda_0\rho^2 T^{-1/2},
\ee
and
\be
\frac{\partial(\rho \mathcal{L})}{\partial \rho}\bigg|_T = (2-\beta) \Lambda_0 \rho T^{1/2}.
\ee
Combining these with Equations \ref{eqn:field} and \ref{eqn:ptot} we have, after more algebra:
\be
\frac{\partial( \mathcal{L}/T)}{\partial T}\bigg|_P = \Lambda_0 \rho T^{-3/2}\bigg[ (1-\delta) - (2-\beta)\frac{\alpha'}{\alpha' + \frac{\partial{\rm ln}P_{\rm rel}}{\partial {\rm ln} \rho}\big|_T}\bigg].
\ee
\noindent
The gas is thus thermally unstable if:
\be
\delta > \frac{(\beta-1)\alpha'+\frac{\partial{\rm ln}P_{\rm rel}}{\partial {\rm ln} \rho}\big|_T}{\alpha'+\frac{\partial{\rm ln}P_{\rm rel}}{\partial {\rm ln} \rho}\big|_T} = \frac{(\beta-1)\alpha'+\gamma}{\alpha'+\gamma} = \frac{(\beta-1)+ \frac{\alpha}{1-\alpha}\gamma}{1+\frac{\alpha}{1-\alpha}\gamma},
\ee
\noindent
where we have assumed that $P_{\rm rel}\propto \rho^\gamma$ and
defined $\alpha\equiv P_{\rm rel} / P_{\rm tot}$.

This equation recovers the result found in \citet{Kim05}.  There they
explored the stability of DF heating, where $\beta=1$, in the absence
of relativistic pressure ($\alpha=0$).  The instability criterion in
this case is $\delta>0$.  As shown in \citet{Kim05}, DF heating
generally yields a temperature dependence such that $\delta>0$, and
thus DF heating alone is thermally unstable.

This equation also implies that heating sources which scale as
$\rho^2$ are thermally stable if the temperature dependence scales as
a power less than $1/2$, which is physically intuitive since $\rho^2
T^{1/2}$ is the scaling of the cooling function.  Moreover, if the
heating source scales as $\rho^2$ then the relativistic fluid has no
influence on the thermal stability.

\end{appendix}

\end{document}